\documentclass{KapProc} 

\usepackage{psfig}

\normallatexbib

\begin{document}

\articletitle[Field Theory Correlators and String Theory]
{Field Theory Correlators and\\ String Theory}


\author{S.S.~Pinsky and U.~Trittmann}
\affil{Department of Physics, Ohio State University\\
Columbus, OH 43210, USA}
\email{pinsky,trittman@mps.ohio-state.edu}
\author{J.R.~Hiller}
\affil{Department of Physics, University of Minnesota-Duluth \\
Duluth, MN 55812, USA}
\email{jhiller@d.umn.edu}

\begin{abstract}
It appears that string-M-theory is the only viable candidate for a
complete theory of matter. It must therefore contain both gravity and QCD.
What is particularly surprising is the recent conjecture that strongly
coupled
QCD matrix elements can be evaluated though a duality with weakly coupled
gravity. To date there has been no direct verification of this conjecture
by Maldacena because of the difficulty of direct strong coupling
calculations 
in gauge theories. We report here on some progress in evaluating a
gauge-invariant correlator in the non-perturbative regime in two and
three dimensions in SYM theories. The calculations are made using
supersymmetric 
discrete light-cone quantization (SDLCQ). We consider a Maldacena-type
conjecture applied to the near horizon  geometry of  a D1-brane in the
supergravity approximation, solve the corresponding ${\cal N}=(8,8)$ SYM
theory in two dimensions, and evaluate the
correlator of the stress-energy tensor. Our numerical results support the
Maldacena conjecture and are within 10-15\%  of the predicted results.
We also present a calculation of the stress-energy correlator in
${\cal N}=1$ SYM theory in 2+1 dimensions. While there is no known duality
relating
this theory to supergravity, the theory does have massless BPS states, and
the
correlator gives important information about the BPS wave function in the
non-perturbative regime.
\end{abstract}


\def\d{\partial}
\def\beq{\begin{equation}}
\def\eeq{\end{equation}}

\section*{Introduction}

Recently the conjecture that certain field theories  admit concrete
realizations
as string theories on particular backgrounds has caused a lot of excitement.
The
original Maldacena conjecture \cite{Maldacena} asserts that the ${\cal N}=4$
supersymmetric Yang-Mills (SYM) theory in 3+1 dimensions is equivalent to
Type
IIB string theory on an $AdS_5\times S^5$ background. However, more
recently,
other string/field theory correspondences have been conjectured. Attempts to
rigorously  test these conjectures have met with only limited success,
because
our understanding of both sides of the correspondences is usually
insufficient. 
The main obstacle is that at the point of correspondence we want  the
curvature
of space-time to be  small in order to use the supergravity
approximation to string theory.   This requires a {\em non-perturbative}
calculation on the field theory side.  We use the method, SDLCQ, in the
corresponding non-perturbative regime. SDLCQ, or Supersymmetric Discretized
Light-Cone Quantization, is a  non-per\-turbative method for solving
bound-state
problems that has been shown to have excellent convergence
properties\cite{lup99}.

Aside from our numerical solutions, there has been very little work on
solving SYM theories using methods that might be described as being from
first
principles.  While selected properties of these theories
have been investigated, one needs the complete solution of the theory to
calculate the correlators. By a ``complete solution'' we mean the spectrum
and the
wave functions of the theory in some well-defined basis. The SYM
theories that are needed for the correspondence with supergravity and string
theory have typically a high degree of supersymmetry and
therefore a large number of fields.  The number of fields significantly
increases
the size of the numerical problem. Therefore, when presenting
the first calculation of
correlators in 2+1 dimensions, we consider only ${\cal N}=1$ SYM.

An important step in these considerations
is to find an observable that can be computed relatively easily
on both sides of a string/field theory correspondence.
It turns out that the correlation function of a gauge invariant operator is
a 
well-behaved object in this sense.
We chose the stress-energy tensor $T^{\mu\nu}$ as this operator
and will construct this observable in the supergravity approximation
to string theory
and perform a non-perturbative SDLCQ calculation of this correlator
on the field theory side.


\section{String Theory}

String theory contains solitons, the so-called D-branes on which modes can
propagate as well as in the bulk. While in general these modes couple, there
exists a limit in which the bulk modes  decouple from the modes on the
D-brane;
this is typically a low energy limit. In this limit the theory on $N$
D$p$-branes, separated by at most  sub-stringy distances becomes a
supersymmetric $SU(N)$ Yang-Mills theory. As the D-branes carry mass and
charge, they can excite gravity modes in the bulk,  for
which in supergravity there exist equivalent solutions. One thus  has a
string/field theory correspondence.  Naively, one would think that
supergravity
can only describe the large distance behavior of fields, but it turns out
that
one can trust these  solutions as long as the curvature is small compared to
the
string scale. In this sense, the large $N$ limit is a valid
description\cite{Itzhaki}.

The most prominent string/field theory correspondence is the so-called
Maldacena conjecture\cite{Maldacena}, which assures that the conformal
${\cal
N}=4$ SYM in 3+1 dimensions, is equivalent to a type IIB string theory on a
$AdS_5\times S^5$ background. In the more general case of non-conformal
theories, it turns out that a black $p$-brane solution, stretching in $p+1$
spacetime dimensions, of supergravity will
correspond to a supersymmetric Yang-Mills theory in $p+1$
dimensions \cite{Itzhaki}.

One can test these string/field theory correspondences,
if one is able to construct and evaluate
observables on both the string and the field
theory regimes. Although this at first seems a hard task, because typically
the small curvature regime of string theory, where the supergravity
approximation allows quantitative calculations, falls into the
strong coupling regime of the field theory side.
We shall see that we can come up with scenarios where we can evaluate the
field theory observable non-perturbatively.

\begin{figure*}
\vspace*{0.5cm}
\centerline{
\unitlength0.8cm
\begin{picture}(15,2)\thicklines
\put(0.2,1){\vector(1,0){14.8}}
\put(0.2,0.8){\line(0,1){0.4}}
\put(5,0.8){\line(0,1){0.4}}
\put(10,0.8){\line(0,1){0.4}}
\put(0,0){$0$}
\put(4.5,0){$\frac{1}{g_{YM}\sqrt{N_c}}$}
\put(9.5,0){$\frac{\sqrt{N_c}}{g_{YM}}$}
\put(14.8,0.2){$x$}
\put(2,0.3){UV}
\put(6.5,0.3){SUGRA}
\put(12,0.3){IR}
\put(1.5,1.5){$N_c^2/x^4$}
\put(6.0,1.5){$N_c^{3/2}/(g_{YM} x^5)$}
\put(11.5,1.5){$N_c/x^4$}
\end{picture}
}
\caption{Phase diagram of two-dimensional ${\cal N}=(8,8)$ SYM:
the theory flows from a CFT in the UV to a conformal $\sigma$-model in the
IR.
The SUGRA approximation is valid in the intermediate range of distances,
$1/g_{YM}\sqrt{N_c}< x < \sqrt{N_c}/g_{YM}$.
\label{figphase}}
\end{figure*}
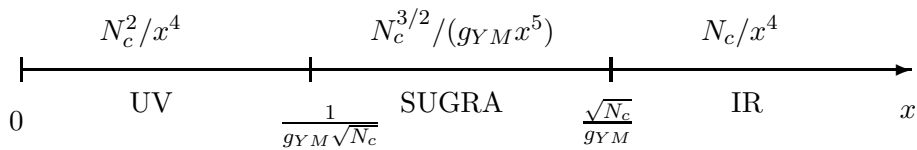

\subsection{Two-dimensional correlation functions from supergravity}

It is instructive to take a closer look on the expected properties
of ${\cal N}=(8,8)$ SYM in two dimensions,
before we proceed to technical details on
the string theory side.
In the extreme ultra-violet (UV) this theory is conformally free and
has a central charge $c_{UV}=N^2_c$. Perturbation theory in turn will be
valid for small effective couplings $g=g_{YM}\sqrt{N_c}x$, where $x$ is
a space coordinate. For large distances, in the far infra-red (IR),
the theory becomes a conformal $\sigma$-model with target space
$(R^8)^{N_c}/S_{N_c}$. The central charge is $c_{IR}=N_c$.
It is a bit more involved to show that here perturbation theory breaks down
when $x{{\sim}} \sqrt{N_c}/g_{YM}$,
see {\em e.g.} Ref.~\cite{Itzhaki}.

The intermediate region, $1/g_{YM}\sqrt{N_c}< x< \sqrt{N_c}/g_{YM}$,
where no perturbative field theoretical description
is possible, is fortunately exactly the region which is accessible to
string theory; or rather, to the supergravity (SUGRA) approximation to
Type IIB string theory on a special background.
It is that of the near horizon geometry of a
D1-brane in the string frame, which has the metric
\begin{eqnarray}
ds^2& =& \alpha' \hat{g}_{YM}
\left( {U^3 \over g^2_s}dx_\parallel^2 +
{dU^2\over U^3} +  U d \Omega_{8-p}^2 \right) \nonumber \\
e^\phi & = & {2 \pi  g_{YM}^2 \over U^3}\hat{g}_{YM},
\label{metric}
\end{eqnarray}
where we defined $\hat{g}_{YM}\equiv 8\pi^{3/2} g_{YM}\sqrt{N_c}$.
In the description of the computation of
the two-point function we follow Ref.~\cite{Anton98b}.
The correlator has been derived in Ref.~\cite{ItzhakiHashimoto}, being
itself
a generalization of Refs.~\cite{Gubser,Witten}.

First, we need to know the action
of the diagonal fluctuations around this background to the quadratic
order. We would like to use the analogue of Ref.~\cite{KRvN}
for our background, Eq.~(\ref{metric}),
which is not (yet) available in the literature.
However, we can identify some diagonal fluctuating degrees of freedom
by following the work on black hole absorption
cross-sections \cite{krasnitz1,krasnitz2}. One can show
that the fluctuations parameterized like
\begin{eqnarray}
ds^2 & = & \left(1 +  f(x^0,U) +  g(x^0,U) \right) g_{00} (dx^0)^2
\nonumber\\
&& +
\left(1 +5 f(x^0,U) +  g(x^0,U)\right) g_{11} (dx^1)^2  \nonumber \\
&& + \left(1 +  f(x^0,U) +  g(x^0,U)\right) g_{UU} dU^2 \nonumber\\
&&+ \left(1 +
f(x^0,U) -    {5 \over 7} g(x^0,U)\right) g_{\Omega\Omega} d \Omega_7^2
\nonumber \\
e^\phi &=& \left(1 + 3 f(x^0,U) - g(x^0,U) \right) e^{\phi_0},
\end{eqnarray}
satisfy the following equations of motion
\begin{eqnarray}\label{fgeq}
f''(U)&=&-{7 \over U}  f'(U) + {g^2_s k^2 \over U^{6}} f(U)
  \\
g''(U) &=&   -{7 \over U} g'(U)+ {72 \over U^2} g(U)  + {g^2_s
k^2 \over U^6} g(U). \nonumber
\end{eqnarray}
Without loss of generality we have assumed here that these
fluctuations vary only along the $x^0$ direction of the world volume
coordinates, and behave like a plane wave. One can interpret a D1-brane
as a black hole in nine dimensions. The fields $f(U)$ and
$g(U)$ are the the minimal set of fixed scalars in
this black hole geometry. In ten dimensions, however, we see that they
are really part of the gravitational fluctuation. Consequently, we expect
that they are associated with the stress-energy tensor in the operator
field correspondence of Refs.~\cite{Gubser,Witten}. In the case of the
correspondence between ${\cal N}{=}4$ SYM field theory and string
theory on an $AdS_5 \times S^5$ background, the
superconformal symmetry allows for the identification of operators and
fields in short multiplets \cite{ferrara}. In the present case of a
D1-brane, we do not have superconformal invariance, and this technique is
not
applicable. Actually, we expect all fields of the theory consistent
with the symmetry of a given operator to mix.  The large distance
behavior should then be dominated by the contribution with the longest
range. The field $f(k^0,U)$ appears to be the one with the longest
range since it is the lightest field.

Eq.~(\ref{fgeq}) for $f(U)$ can be solved explicitly
\begin{equation}
f(U) = U^{-3}  K_{3/2} \left( {\hat{g}_{YM}\over 2 U^{2}} k  \right),
\end{equation}
where $ K_{3/2}(x)$ is a modified Bessel function.
If we take  $f(U)$ to be the analogue of the minimally coupled
scalar, we can construct the flux factor
\begin{eqnarray} 
{\cal F} &=& \lim_{U_0 \rightarrow \infty} {1 \over 2
\kappa_{10}^2} \sqrt{g} g^{UU} e^{-2 (\phi - \phi_{\infty})}\nonumber
\left.\partial_U
\log( f(U))  \right|_{U = U_0} \nonumber\\
&=& {N U_0^2 k^2\over 2 g_{YM}^2} - {N^{3/2}
k^3 \over 4 g_{YM}} + \ldots
\end{eqnarray}
up to a numerical coefficient of order one which we have suppressed.
We see that the leading non-analytic  contribution in $k^2$ is due to
the $k^3$ term.  Fourier transforming the latter yields
\begin{equation}
\langle {\cal O}(x) {\cal O} (0) \rangle = {N^{{3 \over 2}} \over
g_{YM} x^5}. \label{SG}
\end{equation}
This is in line with the discussion at the beginning of this section.
We expect to deviate from the trivial ($1/x^4$)
scaling behavior of the correlator at $x_1={1}/{g_{YM}\sqrt{N_c}}$ and
$x_2={\sqrt{N_c}}/{g_{YM}}$.
This yields the phase diagram in Fig.~\ref{figphase}.
It is interesting to note that
the entire $N_c$ hierarchy is consistent in the sense of Zamolodchikov's
c-theorem, which assures that the central charges obey
$c(x)>c(y)$, whenever $x<y$ \cite{Zamo}.

\subsection{D2-branes and three-dimensional SYM}

As stated in the introduction, in an analogous way one can show that
a system of $D2$-branes corresponds in a certain limit to a
Yang-Mills theory in three dimensions. It is again a ${\cal N}=(8,8)$
supersymmetric theory. Unfortunately, an observable like the correlation
function of the stress-energy tensor has not yet been calculated for this
theory. However, there are encouraging results both on the string and on
the field theory side of the correspondence \cite{Itzhaki,now}.

We describe the phase-diagram of three-dimensional SYM with 16 supercharges
here, following Ref.~\cite{Itzhaki}. Later we will present a
non-perturbative 
field theory calculation within the SDLCQ framework. The latter calculation
is, however, of a theory with an ${\cal N}=1$ supersymmetry. This
theory might 
nevertheless share some features with the full ${\cal N}=(8,8)$ theory,
{\em cf.}~also the results of two-dimensional SYM with different
supersymmetries in Sec.~\ref{formulation}.

It can be argued that the theory has to be described by different degrees
of freedom at different energy scales. At large $N_c$, one can use
perturbation theory of SYM(2+1) in the far ultra-violet, {\em i.e.}
at small distances. The supergravity solution, which in this regime is an
approximation to type IIA string theory with D2-branes, can be trusted at
intermediate distances $r$, $1/g^2_{YM}N_c< r < 1/g^2_{YM}N^{1/5}$.
It has been
conjectured that for large distances, $r>1/g^2_{YM}N^{1/5}$, an M-theory
description is appropriate, while in the far infrared,  $r\gg 1/g^2_{YM}$,
this theory is equivalent to M-theory on an $AdS_4\times S^7$ background,
dual to a CFT with an $SO(8)$ R-symmetry.
This picture is compiled in Fig.~\ref{figphase2}.

\begin{figure*}
\vspace*{0.5cm}
\centerline{
\unitlength0.8cm
\begin{picture}(15,2)\thicklines
\put(0.2,1){\vector(1,0){14.8}}
\put(0.2,0.8){\line(0,1){0.4}}
\put(3.5,0.8){\line(0,1){0.4}}
\put(8,0.8){\line(0,1){0.4}}
\put(11.5,0.8){\line(0,1){0.4}}
\put(0,0){$0$}
\put(2.5,0){$\frac{1}{g^2_{YM}N_c}$}
\put(7,0){$\frac{1}{N_c^{1/5}g^2_{YM}}$}
\put(11,0){$\frac{1}{g^2_{YM}}$}
\put(14.8,0.2){$x$}
\put(1,0.3){SYM$_3$}
\put(4.3,0.3){D2-branes}
\put(8.7,0.3){M2-branes}
\put(12.2,0.3){$AdS_4\times S^7$}
\put(0.5,1.5){perturbative}
\put(4,1.5){IIA string theory}
\put(10.5,1.5){M-theory}
\end{picture}
}
\caption{Phase diagram of three-dimensional ${\cal N}=(8,8)$ SYM:
the theory flows from a perturbative SYM in the UV to a M-theory on
$AdS_4\times S^7$.
The SUGRA approximation is valid in the intermediate range of distances,
$1/g^2_{YM}N_c< x <1/g^2_{YM}N_c^{1/5}$.
\label{figphase2}}
\end{figure*}

\section{Field theory correlators and SDLCQ}
\label{formulation}

Discretized Light-Cone Quantization (DLCQ)
preserves supersymmetry at every stage
of the calculation if the supercharge rather than the Hamiltonian is
diagonalized \cite{Sakai95,hak95}.
The framework of supersymmetric DLCQ (SDLCQ)
allows one to use the advantages of
light-cone quantization ({\em e.g.~}a simpler vacuum)
together with the excellent renormalization properties guaranteed by
supersymmetry.

The technique of (S)DLCQ was reviewed in Ref.~\cite{BPP},
so we can be brief here.  The basic idea of light-cone quantization
is to parameterize space-time using light-cone coordinates
\begin{equation}
x^\pm\equiv \frac{1}{\sqrt{2}}\left(x^0\pm x^1\right),
\end{equation}
and to quantize the theory making $x^+$ play the role of time.
In the discrete light-cone approach, we require the momentum $p_- =
p^+$ along the $x^-$ direction to take on discrete values in units of
$p^+/K$ where $p^+$ is the conserved total momentum of the system.
The integer $K$ is the so-called harmonic resolution, and plays the role
of a discretization parameter.
One can think of this discretization as a consequence of compactifying
the $x^-$ coordinate on a circle with a period $2L = {2 \pi K /
p^+}$. The advantage of discretizing on the light cone is the fact that
the dimension of the Hilbert space becomes finite.  Therefore, the
Hamiltonian is a finite-dimensional matrix, and its dynamics can be
solved explicitly.  In SDLCQ one makes the DLCQ approximation to the
supercharges $Q^i$. Surprisingly, also the discrete representations of $Q_i$
satisfy 
the supersymmetry algebra. Therefore SDLCQ enjoys the improved
renormalization properties of supersymmetric theories.  To
recover the continuum result, $K$ has to go to infinity.
We finds is that SDLCQ usually converges much faster than the
naive DLCQ. 

In the three-dimensional case we also discretize the transverse momentum
along the direction $x^{\perp}$;
however, it is treated in a fundamentally different way.
The transverse resolution is $T$,
and we think of the theory as being compactified on a transverse circle of
length $l$.
Therefore, the transverse momentum is cut off at
$ \pm 2\pi T/l$ and discretized in units of $2\pi/l$. Removal of this
transverse momentum cutoff therefore corresponds to taking the transverse
resolution $T$ to infinity.

Let us now review these ideas in the context of a specific
super-Yang-Mills (SYM) theory.
Actually, it turns out that the two-dimensional SYM is essentially
'included'
in the three-dimensional case \cite{alp99b}, in the sense that in the
weak coupling limit the spectrum of the three-dimension\-al theory is that
of the lower dimensional theory.
We therefore describe here only the 'more general' three-dimensional theory
and hint at the differences and changes that are to make to recover the
two-dimensional theory.
We start with $2+1$ dimensional ${\cal N}=1$ super-Yang-Mills theory
\cite{alp98}
defined on a space-time with one transverse dimension compactified on a
circle.
The action is
\begin{equation}
S=\int d^2 x \int_0^l dx_\perp \mbox{tr}(-\frac{1}{4}F^{\mu\nu}F_{\mu\nu}+
{\rm i}{\bar\Psi}\gamma^\mu D_\mu\Psi).
\end{equation}
After introducing the light--cone coordinates
$x^\pm=\frac{1}{\sqrt{2}}(x^0\pm x^1)$, decomposing the spinor $\Psi$
in terms of chiral projections
\begin{equation}
\psi=\frac{1+\gamma^5}{2^{1/4}}\Psi,\qquad
\chi=\frac{1-\gamma^5}{2^{1/4}}\Psi
\end{equation}
and choosing the light-cone gauge $A^+=0$, we obtain the action
in the form
\begin{eqnarray}\label{action}
S&=&\int dx^+dx^- \int_0^l dx_\perp
\mbox{tr}\left[\frac{1}{2}(\partial_-A^-)^2+
D_+\phi\partial_-\phi+ {\rm i}\psi D_+\psi+ \right.\nonumber \\
& &
\left.
       \hspace{15mm} +{\rm i}\chi\partial_-\chi+\frac{{\rm i}}{\sqrt{2}}\psi
D_\perp\phi+
\frac{{\rm i}}{\sqrt{2}}\phi D_\perp\psi \right].
\end{eqnarray}
A simplification of the
light-cone gauge is that the
non-dynamical fields $A^-$ and $\chi$ may be explicitly
solved from their Euler--Lagrange equations of motion
\begin{equation}
A^-=\frac{g_{\rm YM}}{\partial_-^2}J=
\frac{g_{\rm 
YM}}{\partial_-^2}\left(i[\phi,\partial_-\phi]+2\psi\psi\right), \quad
\chi=-\frac{1}{\sqrt{2}\partial_-}D_\perp\psi.\nonumber
\end{equation}

These expressions may be used to express any operator
in terms of the physical degrees of freedom only.
In particular, the light-cone energy, $P^-$, and momentum
operators, $P^+$,$P^{\perp}$,
corresponding to  translation
invariance in each of the coordinates
$x^\pm$ and $x_\perp$ may be calculated explicitly as
\begin{eqnarray}\label{moment}
P^+&=&\int dx^-\int_0^l dx_\perp\mbox{tr}\left[(\partial_-\phi)^2+
{\rm i}\psi\partial_-\psi\right],\\
P^-&=&\int dx^-\int_0^l dx_\perp\mbox{tr}
\left[-\frac{g_{\rm YM}^2}{2}J\frac{1}{\partial_-^2}J-
          \frac{{\rm
i}}{2}D_\perp\psi\frac{1}{\partial_-}D_\perp\psi\right],\\
P_\perp &=&\int dx^-\int_0^l
dx_\perp\mbox{tr}\left[\partial_-\phi\partial_\perp\phi+
          {\rm i}\psi\partial_\perp\psi\right].
\end{eqnarray}
The light-cone supercharge in this theory
is a two-component Majorana spinor, and may be conveniently
decomposed in terms of its chiral projections
\begin{eqnarray}\label{sucharge}
Q^+&=&2^{1/4}\int dx^-\int_0^l
dx_\perp\mbox{tr}\left[\phi\partial_-\psi-\psi\partial_-
                 \phi\right],\\
Q^-&=&2^{3/4}\int dx^-\int_0^l
dx_\perp\mbox{tr}\left[2\partial_\perp\phi\psi+
          g_{\rm YM}\left({\rm
i}[\phi,\partial_-\phi]+2\psi\psi\right)\frac{1}{\partial_-}\psi\right].
\nonumber
\end{eqnarray}
The action (\ref{action}) gives the following canonical
(anti-)commutation relations for
propagating fields for large $N_c$ at equal $x^+$:
\begin{eqnarray}
\label{comm}
\left[\phi_{ij}(x^-,x_\perp),\partial_-\phi_{kl}(y^-,y_\perp)\right]&=&
\left\{\psi_{ij}(x^-,x_\perp),\psi_{kl}(y^-,y_\perp)\right\}\\
&=&
\frac{1}{2}\delta(x^- -y^-)\delta(x_\perp -y_\perp)\delta_{il}\delta_{jk}.
\nonumber
\end{eqnarray}
Using these relations one can check the supersymmetry algebra
\begin{equation}
\{Q^\pm,Q^\pm+\}=2\sqrt{2}P^\pm,\qquad
\{Q^+,Q^-\}=-4P_\perp.
\label{superr}
\end{equation}

In solving for mass eigenstates, we will consider
only states which have vanishing transverse momentum,
which is possible since the total transverse momentum operator
is kinemat\-ical. Strictly speaking, on a transverse
cylinder, there are separate sectors with total
transverse momenta $2\pi N_\perp/L$; we consider only one of them,
$N_\perp=0$.
On such states, the light-cone supercharges
$Q^+$ and $Q^-$ anti-commute with each other, and the supersymmetry algebra
is equivalent to the ${\cal N}=(1,1)$ supersymmetry
of the dimensionally reduced ({\em i.e.}, two-dimensional) theory
\cite{Sakai95}.
Moreover, in the $P_{\perp} = 0$ sector,
the mass squared operator $M^2$ is given by
$M^2=2P^+P^-$.

As we mentioned earlier, in order to render the bound-state equations
numerically tract\-able, the transverse
momenta of partons must be truncated.
First, we introduce the Fourier expansion for the fields $\phi$ and $\psi$,
where the transverse space-time coordinate $x_{\perp}$ is periodically
identified
\begin{eqnarray}
&&\phi_{ij}(0,x^-,x_\perp) =
\frac{1}{\sqrt{2\pi l}}\sum_{n^{\perp} = -\infty}^{\infty}
\int_0^\infty
       \frac{dk^+}{\sqrt{2k^+}}\\
&&
\quad\times\left[
       a_{ij}(k^+,n^{\perp})e^{-{\rm i}k^+x^- -{\rm i}
\frac{2 \pi n^{\perp}}{l} x_\perp}+
       a^\dagger_{ji}(k^+,n^{\perp})e^{{\rm i}k^+x^- +
{\rm i}\frac{2 \pi n^{\perp}}{l}  x_\perp}\right]\,,
\nonumber\\
&&\psi_{ij}(0,x^-,x_\perp) =
\frac{1}{2\sqrt{\pi l}}\sum_{n^{\perp}=-\infty}^{\infty}\int_0^\infty
       dk^+ 
\\&&
\quad\times\left[b_{ij}(k^+,n^{\perp})e^{-{\rm i}k^+x^- -
{\rm i}\frac{2 \pi n^{\perp}}{l} x_\perp}+
       b^\dagger_{ji}(k^+,n^\perp)e^{{\rm i}k^+x^- +{\rm i}
\frac{2 \pi n^{\perp}}{l} x_\perp}\right]\,.
\nonumber
\end{eqnarray}
Substituting these into the (anti-)commutators (\ref{comm}),
one finds
\begin{eqnarray*}
\left[a_{ij}(p^+,n_\perp),a^\dagger_{lk}(q^+,m_\perp)\right]&=&
\left\{b_{ij}(p^+,n_\perp),b^\dagger_{lk}(q^+,m_\perp)\right\}\\
&=&\delta(p^+ -q^+)\delta_{n_\perp,m_\perp}\delta_{il}\delta_{jk}.
\end{eqnarray*}
The supercharges then take the following form:
\begin{eqnarray}\label{TruncSch}
&&Q^+={\rm i}2^{1/4}\sum_{n^{\perp}\in {\bf Z}}\int_0^\infty dk\sqrt{k}\\
&&\quad\quad\quad\quad\quad\quad\times\left[
b_{ij}^\dagger(k,n^\perp) a_{ij}(k,n^\perp)-
a_{ij}^\dagger(k,n^\perp) b_{ij}(k,n^\perp)\right],\nonumber\\
\label{Qminus1}
&&Q^-=\frac{2^{7/4}\pi {\rm i}}{l}\sum_{n^{\perp}\in {\bf Z}}\int_0^\infty
dk
\frac{n^{\perp}}{\sqrt{k}}\\&&
\qquad\quad\quad\quad\quad\quad\times\left[
a_{ij}^\dagger(k,n^\perp) b_{ij}(k,n^\perp)-
b_{ij}^\dagger(k,n^\perp) a_{ij}(k,n^\perp)\right]+\nonumber\\
&&+ {{\rm i} 2^{-1/4} {g_{\rm YM}} \over \sqrt{l\pi}}
\sum_{n^{\perp}_{i} \in {\bf Z}} \int_0^\infty dk_1dk_2dk_3
\delta(k_1+k_2-k_3) \delta_{n^\perp_1+n^\perp_2,n^\perp_3}
\left\{ \frac{}{} \right.\nonumber\\
&&{1 \over 2\sqrt{k_1 k_2}} {k_2-k_1 \over k_3}
[a_{ik}^\dagger(k_1,n^\perp_1) a_{kj}^\dagger(k_2,n^\perp_2)
b_{ij}(k_3,n^\perp_3)\nonumber\\
&&\qquad\qquad\qquad\qquad\qquad\qquad
-b_{ij}^\dagger(k_3,n^\perp_3)a_{ik}(k_1,n^\perp_1)
a_{kj}(k_2,n^\perp_2) ]\nonumber\\
&&{1 \over 2\sqrt{k_1 k_3}} {k_1+k_3 \over k_2}
[a_{ik}^\dagger(k_3,n^\perp_3) a_{kj}(k_1,n^\perp_1) b_{ij}(k_2,n^\perp_2)
\nonumber\\
&&\qquad\qquad\qquad\qquad\qquad\qquad
-a_{ik}^\dagger(k_1,n^\perp_1) b_{kj}^\dagger(k_2,n^\perp_2)
a_{ij}(k_3,n^\perp_3) ]\nonumber\\
&&{1 \over 2\sqrt{k_2 k_3}} {k_2+k_3 \over k_1}
[b_{ik}^\dagger(k_1,n^\perp_1) a_{kj}^\dagger(k_2,n^\perp_2)
a_{ij}(k_3,n^\perp_3)
\nonumber\\
&&\qquad\qquad\qquad\qquad\qquad\qquad
-a_{ij}^\dagger(k_3,n^\perp_3)b_{ik}(k_1) a_{kj}(k_2,n^\perp_2) ]\nonumber\\
&& ({ 1\over k_1}+{1 \over k_2}-{1\over k_3})
[b_{ik}^\dagger(k_1,n^\perp_1) b_{kj}^\dagger(k_2,n^\perp_2)
b_{ij}(k_3,n^\perp_3)
\nonumber\\
&&
\qquad\qquad\qquad\qquad\qquad\qquad
\left.+b_{ij}^\dagger(k_3,n^\perp_3) b_{ik}(k_1,n^\perp_1)
b_{kj}(k_2,n^\perp_2)]
       \frac{}{}\right\}. \nonumber
\end{eqnarray}
We now perform the truncation procedure; namely, in all sums over the
transverse momenta $n^{\perp}_{i}$ appearing in the above expressions for
the
supercharges, we restrict summation to the following allowed momentum
modes: $n^{\perp}_{i}=0,\pm 1 ... \pm T$.  Note that this prescription is
symmetric, in the sense that there are as many positive modes as there are
negative ones. In this way we  retain parity symmetry in the transverse
direction. The longitudinal momenta $k_i=n_i \pi/L$ are restricted by the
longitudinal resolution according to $K=\sum_i n_i$.

The two-dimensional supercharges are essentially recovered, when we put
$n_{\perp}$ to zero. In particular, the first term of the supercharge $Q^-$,
Eq.~(\ref{Qminus1}), is absent in this case. Additionally,
we have to adjust the
normalization constants in front of the expressions for the supercharges.

\subsection{Two dimensional correlators}

Using SDLCQ, we can reproduce the SUGRA scaling relation, Eq.~(\ref{SG}),
fix the numerical coefficient, and calculate the
cross-over behavior at $1/g_{YM}\sqrt{N_c}<x<\sqrt{N_c}/g_{YM}$.
To exclude subtleties, {\em nota bene} issues of zero modes,
we checked  our results
against the free fermion and the 't Hooft model and found consistent
results.

Let us now focus on the theory in two dimensions.
We would like to compute a
general expression for the correlator of the form
$F(x^-,x^+) = \langle {\cal O}(x^-,x^+) {\cal O} (0,0) \rangle$\@.
In DLCQ one fixes the total momentum in the $x^-$ direction, and it is
natural to compute the Fourier transform and express it
in a spectrally decomposed form
\begin{eqnarray}
\tilde{F}(P_-,x^+) &=& {1 \over 2 L} \langle {\cal O}(P_-,x^+) {\cal
O}(-P_-,
0) \rangle\nonumber \\
&=&\sum_n {1 \over 2 L} \langle 0| {\cal O}(P_-) | n
\rangle e^{-i P_+^n x^+} \langle n|  {\cal O}(-P_-,0) |0 \rangle\ .
\end{eqnarray}
The form of the correlation function in position space is
then recovered by Fourier
transforming with respect to $P_- =K\pi/L$.  We can continue to Euclidean
space by taking $r = \sqrt{2 x^+ x^-}$ to be real. The result for the
correlator of the  stress-energy tensor is
\begin{eqnarray}
F(x^-,x^+)&=&
\sum_n\left|{L \over \pi} \langle n | T^{++}(-K) |0 \rangle \right|^2
\left({x^+ \over x^-}\right)^2\nonumber\\
&&\times {M_n^4 \over 8 \pi^2 K^3}
K_4\left(M_n\sqrt{2 x^+ x^-}\right),\label{master}
\end{eqnarray}
where $M_i$ is a mass eigenvalue and $K_4(x)$ is the
modified Bessel function of order 4.
Note that this quantity depends on the harmonic resolution $K$,
but involves no other unphysical quantities. In particular,
the expression is independent of the box length $L$.

The matrix element $(L/\pi) \langle 0 | T^{++}(K) | i \rangle$
can be substituted directly to give an
explicit expression for the two-point function. We see immediately that
the correlator has the correct small-$r$ behavior, for in that limit, it
asymptotes to
\[
\left({x^- \over x^+ }\right)^2 F(x^-,x^+) =
   {N_c^2(2 n_b + n_f) \over 4  \pi^2 r^4}
            \left(1 - {1 \over K}\right),
\]
which we expect for the theory of $n_b(n_f)$ free bosons (fermions)
at large $K$.
On the other hand, the contribution to the correlator from strictly massless
states is given by
\begin{eqnarray}
\left( x^- \over x^+ \right)^2 F(x^-,x^+)=    {6 \over K^3 \pi^2 r^4}
\sum_i\left| {L \over \pi}
            \langle 0 | T^{++}(K) | i\rangle \right|^2_{M_i=0}. \nonumber
\end{eqnarray}
It is important to notice
that this $1/r^4$ behavior at large $r$ is {\em not} the one
we are looking for at large $r$. First of all, we do not
expect any massless physical bound state in this theory, and,
additionally, it has the wrong $N_c$ dependence.
Relative to the $1/r^4$ behavior at small $r$, the $1/r^4$ behavior at
large $r$ that we expect is down by a factor of $1/N_c$. This behavior
is suppressed because we are
performing a large-$N_c$ calculation. All we can
hope is to see the transition from the $1/r^4$ behavior at small $r$ to the
region where the correlator behaves like $1/r^5$.

\subsubsection{The ${\cal N}=(1,1)$ theory}
Although it is the ${\cal N}=(8,8)$ theory  in which we
are ultimately interested in, we can, nevertheless,
perform the computation of the correlation function in
models with less supersymmetry.
The evaluation of the correlator for the stress energy tensor in the
${\cal N}=(8,8)$ theory is especially hard because of the many degrees
of freedom due to the large number of supercharges in that theory.
We will show in Sec.~\ref{Sec} how to overcome this obstacle by exploiting
a residual 'flavor' symmetry of the theory.
To see how our numerical method works without these complications, it might
be worthwhile to study the theory with supercharges (1,1).
In the next section we will briefly cover the theory with
a ${\cal N}=(2,2)$ supersymmetry.

It has been argued that the
${\cal N}=(1,1)$ SYM theory does not exhibit dynamical supersymmetry
breaking.
A physicist's proof that supersymmetry is not spontaneously broken in this
theory was given in Ref.~\cite{Anton98b}.
This theory is also believed not to be confining \cite{adi1}\cite{adi2}, and
is therefore expected to exhibit non-trivial infra-red dynamics.
The SDLCQ of the 1+1 dimensional model with ${\cal N}=(1,1)$ supersymmetry
was solved in Refs.~\cite{Sakai95,alp98a}, and we apply these results
directly in order to compute (\ref{master}).  For simplicity, we
work at large $N_c$.
The spectrum of this theory at finite $K$
consists of $2K-2$ exactly massless states,
{\em i.e.} $K-1$ massless bosons, and their superpartners, accompanied
by large numbers of massive states separated by a gap. There is
numerical evidence that this gap closes in the continuum limit.
At finite $N_c$, we expect the degeneracy of $2K-2$ exactly
massless states to be broken, giving rise to precisely a continuum
of states starting at $M=0$ as expected.

\begin{figure}
\centerline{
\psfig{file=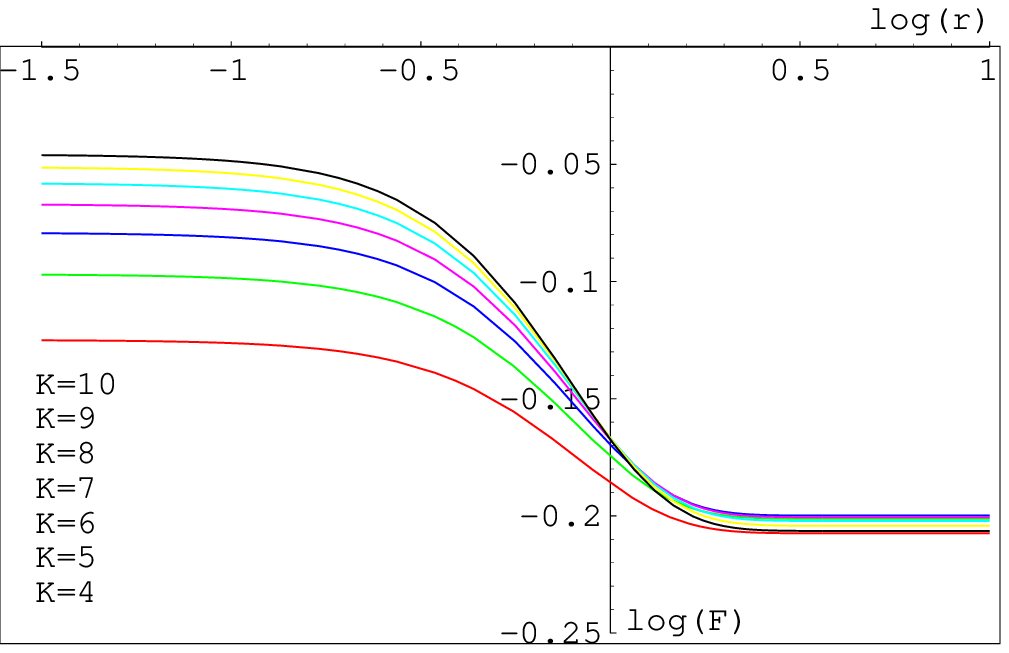,width=5.85cm}
\psfig{file=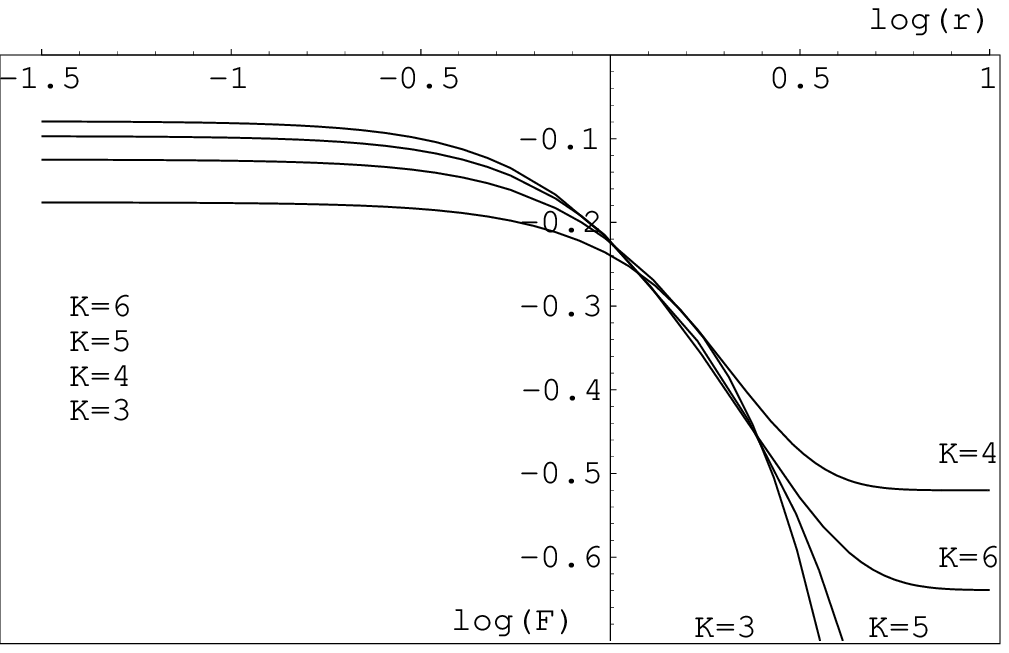,width=6cm}}
\caption{Log-Log
plots of the two-dimensional
correlation function $\langle T^{++}(x) T^{++}(0) \rangle
\left({x^- \over x^+} \right)^2 {4 \pi^2 r^4 \over N^2 (2 n_b +n_f)}$
v.s. $r$ at $g_{YM}^2 N /\pi = 1$.
Left: (a) ${\cal N}=(1,1)$ SYM for $K=4$ to $10$.
Right: (b) ${\cal N}=(2,2)$ SYM for $K=3,4,5,6$.
\label{N12}}
\end{figure}

The stress-energy correlator of this theory for
various values of the harmonic resolution $K$, is shown in Fig.~\ref{N12}(a).
We find the curious feature that it asymptotes to the inverse
power law $c/r^4$ for large $r$. This behavior comes about due to the
coupling $\langle 0 | T^{++} | n \rangle$ with exactly massless states
$|n \rangle$. The contribution to (\ref{master}) from strictly massless
states are given by
\begin{eqnarray}
\left( x^- \over x^+ \right)^2 F(x^-,x^+) &=& \left. \left| {L \over \pi}
\langle 0 | T^{++}(k) | n \rangle \right|^2 {M_n^4 \over 8 \pi^2 k^3 }
K_4(M_n r) \right|_{M_n =0}\\
&=&\left| {L \over \pi} \langle 0 | T^{++}(k) | n \rangle \right|^2_{M_n=0}
{6 \over k^3 \pi^2 r^4} .
\end{eqnarray}
We have computed this quantity as a function of the inverse
harmonic resolution $1/K$ and extrapolated to the continuum limit.
The data currently available
suggests that the non-zero contribution from these massless states
persists in this limit.

%
%

\subsubsection{The ${\cal N}=(2,2)$ theory}

Let us now turn to the model with ${\cal N}=(2,2)$  supersymmetry. The SDLCQ
version of this model was solved in Ref.~\cite{dlcq22}. The result of this
computation can be inserted into Eq.~(\ref{master}). The result is
shown in Fig.~\ref{N12}(b). This model appears to exhibit the
onset of a gapless continuum of states more rapidly than the ${\cal
N}=(1,1)$
model as the harmonic resolution $K$ is increased. Just as we found in
the latter model, this theory contains exactly massless states in the
spectrum.  These massless states appear to couple to $T^{++} | 0
\rangle$ only for even $K$, and the overlap appears to be decreasing
for growing $K$. It is believed that this model is likely to exhibit a
power law behavior $c/r^\gamma$ for $\gamma > 4$ for the $T^{++}$
correlator for $r \gg g_{YM} \sqrt{N}$ in the large $N_c$
limit \cite{Anton98b}.

\subsubsection{The  ${\cal N}=(8,8)$ theory}
\label{Sec}

In principle, we can now
calculate the correlator numerically by evaluating Eq.~(\ref{master}).
However, it turns out that even for very modest harmonic resolutions,
we face a tremendous numerical task.
At $K=2,3,4$, the dimension of the associated Fock space is
$256, 1632,$ and $29056$, respectively.
The usual procedure is to
diagonalize the Hamiltonian $P^-$ and then to evaluate the projection
of each eigenfunction on the fundamental state $T^{++}(-K)|0\rangle$.
Since we are only interested in states which
have nonzero value of such projection, we are able to significantly
reduce our numerical efforts.

In the continuum limit, the result does not depend on which of the eight
supercharges $Q^-_{\alpha}$ one
chooses. In DLCQ, however,
the situation is a bit subtler:
while the spectrum of $(Q_\alpha^-)^2$ is the same for all
$\alpha$, the wave functions depend on the choice of supercharge
\cite{Anton98}. This dependence is an
artifact of the discretization and disappears in the continuum limit.
What happens if we just pick one supercharge, say $Q^-_1$?
Since the state $T^{++}(-K)|0\rangle$
is a singlet under R--symmetry acting on the ``flavor'' index of
$Q^-_{\alpha}$, the correlator (\ref{master}) does not depend on the choice
of
$\alpha$ even at finite resolution!

We can exploit this fact to simplify our calculations.
Consider 
an operator $S$ commuting with both $P^-$ and $T^{++}(-K)$, and such that
$S|0 \rangle=s_0|0 \rangle$. Then the Hamiltonian and $S$ can be
diagonalized
simultaneously.  We assume in the sequel that the set of states
$|i \rangle$ is a result of
such a diagonalization. In this case, only states satisfying the condition
$S|i \rangle=s_0|i \rangle$ contribute to the sum in (\ref{master}), and we
only need to diagonalize $P^-$ in this sector, which reduces
the size of the problem immensely.
We can deduce from the structure of the state
$T^{++}(-K)|0\rangle$ that any transformation of the form
\begin{eqnarray}
a^I_{ij}(k)\rightarrow f(I)a^{P[I]}_{ij}(k), \qquad f(I)=\pm 1
 \nonumber  \\ 
b^\alpha_{ij}(k)\rightarrow g(\alpha)b^{Q[\alpha]}_{ij}(k), \qquad
g(\alpha)=\pm 1 
\label{initsymm}
\end{eqnarray} 
given arbitrary permutations $P$ and $Q$ of the $8$ flavor indices,
commutes with $T^{++}(-K)$.
The vacuum will then be an eigenstate of
this transformation with eigenvalue $1$. The requirement for $P^-=(Q^-_1)^2$
to be invariant under $S$ imposes some restrictions on the permutations.
In fact, we will require that $Q^-_1$ be invariant under $S$, in order
to guarantee that $P^-$ is invariant.

The form of the supercharge from \cite{Anton98} is
\begin{eqnarray}\label{Qminus}
&&Q^-_{\alpha} =  \int_0^{\infty}
  [...]b^{\dagger}_{\alpha}(k_3)a_{I}(k_1)a_{I}(k_2) +...\\
&&\!\!\!\!\!\!\!\!+(\beta_I \beta_J^T - \beta_J \beta_I^T )_{\alpha \beta}
  [..]  b^{\dagger}_{\beta}(k_3)a_{I}(k_1)a_{J}(k_2) + \ldots .\nonumber
\end{eqnarray}
Here the $\beta_I$ are $8\times 8$ real matrices satisfying
$\{\beta_I,\beta_J^T \} = 2\delta_{IJ}$.

Let us consider the expression for $Q^-_1$, Eq.~(\ref{Qminus}).
The first part of the supercharge
does not include $\beta$ matrices, and is therefore invariant under
the transformation, Eq.~(\ref{initsymm}),
as long as $g(1)=1$ and $Q[1]=1$. We will consider only such
transformations.
The crucial observation for the analysis
of the symmetries of the $\beta$ terms is
that in the representation of the $\beta$ matrices we have chosen,
the expression ${\cal B}^\alpha_{IJ}=
\left(\beta_I\beta^T_J-\beta_J\beta^T_I\right)_{1\alpha} $
may take only the values $\pm 2$ or zero. Besides, for any pair
$(I,J)$ there is only one (or no)
value of $\alpha$ corresponding to nonzero ${\cal B}$. Using this
information,
we may represent ${\cal B}$ in a compact form. With
the definition \cite{this work}
{\small
\begin{equation}
\mu_{IJ}=\left\{\begin{array}{rl} \alpha\,, & {\cal B}^\alpha_{IJ}=2 \\
                                  -\alpha\,, & {\cal B}^\alpha_{IJ}=-2 \\
                                     0\,,    & {\cal B}^\alpha_{IJ}=0
                                                {\mbox{ for all }}\alpha
                 \end{array}\right.,
\end{equation}
}
together with the special choice of $\beta$ matrices we get
the following expression for $\mu$
{\small
\[
\mu=\left(
\begin{array}{rrrrrrrr}
0&5&-7&2&-6&3&-4&8\\
-5&0&-3&6&2&-7&8&4\\
7&3&0&-8&-4&-5&6&2\\
-2&-6&8&0&-5&4&3&7\\
6&-2&4&5&0&-8&-7&3\\
-3&7&5&-4&8&0&-2&6\\
4&-8&-6&-3&7&2&0&5\\
-8&-4&-2&-7&-3&-6&-5&0\\
\end{array}
\right).
\]
}
The next step is to look for a subset of the transformations,
Eq.~(\ref{initsymm}), which satisfy
the conditions $g(1)=1$ and $Q[1]=1$ and leave the matrix $\mu$ invariant.
This invariance implies that
\begin{equation}
Q[\mu_{P[I]P[J]}]=g(\mu_{IJ})f(I)f(J)\mu_{IJ}\,.
\label{SymmFinalCond}
\end{equation}
The subset of transformations we are looking for forms a subgroup $R$ of
the permutation group $S_8\times S_8$. Consequently, we will search
for the elements of $R$ that
square to one. Products of such elements generate the whole group
in the case of $S_8\times S_8$.
We will show later that this remains true for $R$.
Not all of the $Z_2$ symmetries satisfying (\ref{SymmFinalCond})
are independent. In particular, if $a$ and $b$ are two such
symmetries then $aba$ is also a valid $Z_2$ symmetry. By going
systematically 
through the different
possibilities, we have found that there are $7$ independent $Z_2$ symmetries
in
the group $R$. They are listed in Table \ref{tab1}.
We explicitly constructed all the symmetries of the type,
Eq.~(\ref{initsymm}),
which satisfy Eq.~(\ref{SymmFinalCond}) using {\sc Mathematica}.
It turns out that the group
of such transformations has $168$ elements, and we have shown that all
of them can be generated from the seven $Z_2$ symmetries mentioned above.
{\footnotesize
\begin{table*}
\centerline{
\begin{tabular}{|c|c|c|c|c|c|c|c|c|c|c|c|c|c|c|c|c|}
\hline
   &$a_1$&$a_2$&$a_3$&$a_4$&$a_5$&$a_6$&$a_7$&$a_8$&$b_2$&$b_3$&$b_4$&$b_5$&
   $b_6$&$b_7$&$b_8$\\
\hline
1 
&$a_7$&$a_3$&$a_2$&$a_6$&$a_8$&$a_4$&$a_1$&$a_5$&$b_2$&$-b_3$&$-b_4$&$-b_6$&
   $-b_5$&$b_8$&$b_7$\\
\hline
2  
&$a_3$&$a_6$&$a_1$&$a_5$&$a_4$&$a_2$&$a_8$&$a_7$&$-b_4$&$b_3$&$-b_2$&$-b_5$&
   $b_8$&$-b_7$&$b_6$\\
\hline
3  
&$a_8$&$a_7$&$a_6$&$a_5$&$a_4$&$a_3$&$a_2$&$a_1$&$-b_3$&$-b_2$&$b_4$&$-b_5$&
   $b_7$&$b_6$&$-b_8$\\
\hline
4  
&$a_5$&$a_4$&$a_8$&$a_2$&$a_1$&$a_7$&$a_6$&$a_3$&$-b_2$&$-b_7$&$b_8$&$b_5$&
   $-b_6$&$-b_3$&$b_4$\\
\hline
5  
&$a_8$&$a_3$&$a_2$&$a_7$&$a_6$&$a_5$&$a_4$&$a_1$&$-b_5$&$-b_3$&$b_7$&$-b_2$&
   $b_6$&$b_4$&$-b_8$\\
\hline
6  
&$a_5$&$a_8$&$a_7$&$a_6$&$a_1$&$a_4$&$a_3$&$a_2$&$-b_8$&$b_5$&$-b_4$&$b_3$&
   $-b_6$&$b_7$&$-b_2$\\
\hline
7  
&$a_4$&$a_6$&$a_8$&$a_1$&$a_7$&$a_2$&$a_5$&$a_3$&$-b_2$&$-b_6$&$b_5$&$b_4$&
   $-b_3$&$-b_7$&$b_8$\\
\hline
\end{tabular}}
\caption{Seven independent $Z_2$
symmetries of the group $R$,
which act on the 'flavor' quantum number of the different particles.
Under the first of these symmetries, {\em e.g.}, the boson $a_1$ is
transformed
into $a_7$, etc.}
\label{tab1}
\end{table*}}

In our numerical algorithm we implemented the $Z_2$ symmetries as follows.
We can group the Fock states in classes and treat the whole class
as a new state, because all states relevant for the correlator are
singlets under the symmetry group $R$.
As an example, consider the simplest non-trivial singlet
\begin{equation}
|1\rangle=\frac{1}{8}\sum_{I=1}^8 {\mbox{tr}}
\left(a^\dagger(1,I)a^\dagger(K-1,I)\right)|0\rangle.
\end{equation}
Hence, if we encounter the state
$a^\dagger(1,1)a^\dagger(K-1,1)|0\rangle$
while constructing the basis, we will replace it by the class
representative; in this case, by the state $|1\rangle$. Such a procedure
significantly decreases the size of the basis, while keeping all the
information necessary for calculating the correlator.
In summary, this use of the
discrete flavor symmetry of the problem reduces the size
of the Fock space by orders of magnitude.

In addition to these simplifications, one can further improve on the
numerical 
efficiency by using Lanczos diagonalization techniques\cite{Lanczos}.
Namely, we substitute the explicit diagonalization with an efficient
approximation.
The idea is to use a symmetry preserving (Lanczos) algorithm.
If we start with a normalized vector $|u_1\rangle$ proportional to
the fundamental state $T^{++}(-K)|0\rangle$,
the Lanczos recursion will produce a
tridiagonal representation of the Hamiltonian
$H_{LC}=2P^+P^-$. 
Due to orthogonality of $\{|u_i\rangle\}$,
only the (1,1) element of the exponential of the tridiagonal matrix
$\hat{H}_{LC}$ will contribute to the correlator \cite{Haydock}.
We exponentiate by diagonalizing $\hat{H}_{LC}\vec{v}_i=\lambda_i\vec{v}_i$
with eigenvalues $\lambda_i$ and get
\[
F(P^+,x^+)=\frac{|N_0|^{-2}}{2L}\left(\frac{\pi}{L}\right)^2
\sum_{j=1}^{N_L}|(v_j)_1|^2e^{-i\frac{\lambda_j L}{2K\pi}x^+}.
\] 
Finally, we Fourier transform to obtain
\begin{eqnarray*}
F(x^-,x^+)&=&\frac{1}{8\pi^2K^3}\left(\frac{x^+}{x^-}\right)^2\frac{1}{|N_0|
^2}\sum_{j=1}^{N_L}|(v_j)_1|^2 \lambda_j^2 K_4(\sqrt{2x^+x^-\lambda_i}),
\nonumber
\end{eqnarray*}
which is equivalent to Eq.~(\ref{master}).
This algorithm is correct
only if the number of Lanczos iterations $N_L$ runs up to the rank
of the original matrix, but {\em in praxi}
already a basis of about 20 vectors covers all leading contributions
to the correlator.

\begin{figure}
\centerline{
\psfig{file=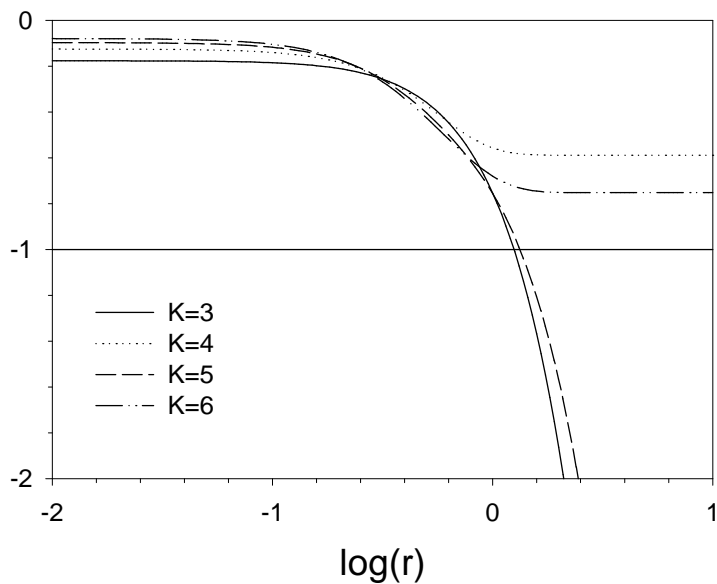,width=10cm}}
\centerline{(a)}
\centerline{
\psfig{file=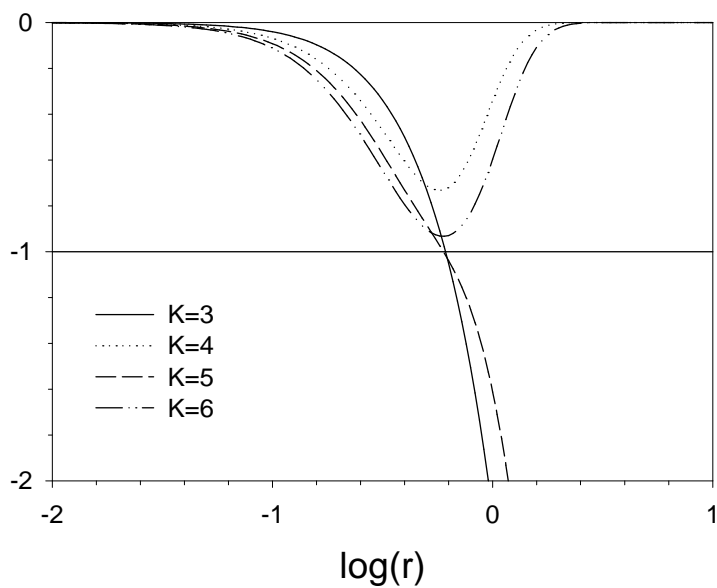,width=10cm}}
\centerline{(b)}
\caption{Top: (a) Log-Log
plot of ${\cal F}(r)=\langle T^{++}(x) T^{++}(0) \rangle
\left({x^- \over x^+} \right)^2 {4 \pi^2 r^4 \over N_c^2 (2 n_b +n_f)}$
vs. $r$ for $g_{YM}^2 N_c /\pi = 1.0$, $K=3,4,5$ and
$6$. Bottom: (b) the log-log
derivative with respect to $r$ of the correlation function in (a).
\label{fig}}
\end{figure}


{\bf Numerical Results.}
To evaluate the expression for the correlator
${\cal F}(r)$, we have to calculate the mass spectrum and insert it into
Eq.~(\ref{master}).
In the ${\cal N}=(8,8)$ supersymmetric Yang-Mills
theory the contribution of massless states becomes a problem.
These states exist in the SDLCQ calculation, but are unphysical.
It has be shown that theses states are not normalizable
and that the number of partons in these states is even (odd) for
$K$ even (odd) \cite{Anton98}.
Because the correlator is only sensitive to two particle contributions,
the curves ${\cal F}(r)$ are different for even and odd $K$.
Unfortunately, the 
unphysical states yield also the typical $1/r^4$ behavior,
but have a wrong $N_c$ dependence.
The regular $1/r^4$ contribution is down by $1/N_c$, so we cannot
see this contribution at large $r$, because we are working in the large
$N_c$ limit.

We can use this information about the unphysical states, however,
to determine when our approximation breaks down.
It is the region where the unphysical massless states dominate the
correlator sum.  Unfortunately, this is also the region where we
expect the true large-$r$ behavior to dominate the correlator, if
only the extra states were absent.
In Fig.~\ref{fig}(a) for even resolution, the region where the
correlator  starts to behave like
$1/r^4$ at large $r$
is clearly visible. In Fig.~\ref{fig}(b) we see that for even
resolution the effect of the massless state on the derivative
is felt at smaller values
of $r$ where the even resolution curves start to turn up.
Another estimate of where this
approximation breaks down, that gives consistent values, is the set of
points where the even and odd resolution derivative curves cross. We do
not expect these curves to cross on general grounds, based on work
in \cite{Anton98b}, where we considered a number of other theories.
Our calculation is consistent in the sense that this breakdown
occurs at larger and larger $r$ as $K$ grows.

We expect to approach the line $d{{\cal F}(r)}/dr=-1$ line
signaling the cross-over from the trivial $1/r^4$ behavior
to the characteristic $1/r^5$ behavior of the supergravity correlator,
Eq.~(\ref{SG}). Indeed, the derivative curves in Fig.~\ref{fig}(b)
are approaching $-1$ as we increase the
resolution and appear to be about $85-90\%$ of this value
before the approximation
breaks down. There is, however, no indication of convergence yet; therefore,
we cannot claim a numerical proof of the Maldacena conjecture.
A safe signature of equivalence of the field and string theories
would be if the derivative curve
would flatten out at $-1$ before the approximation breaks down.


\subsection{Three-dimensional correlators}

\begin{figure}
\centerline{
\psfig{file=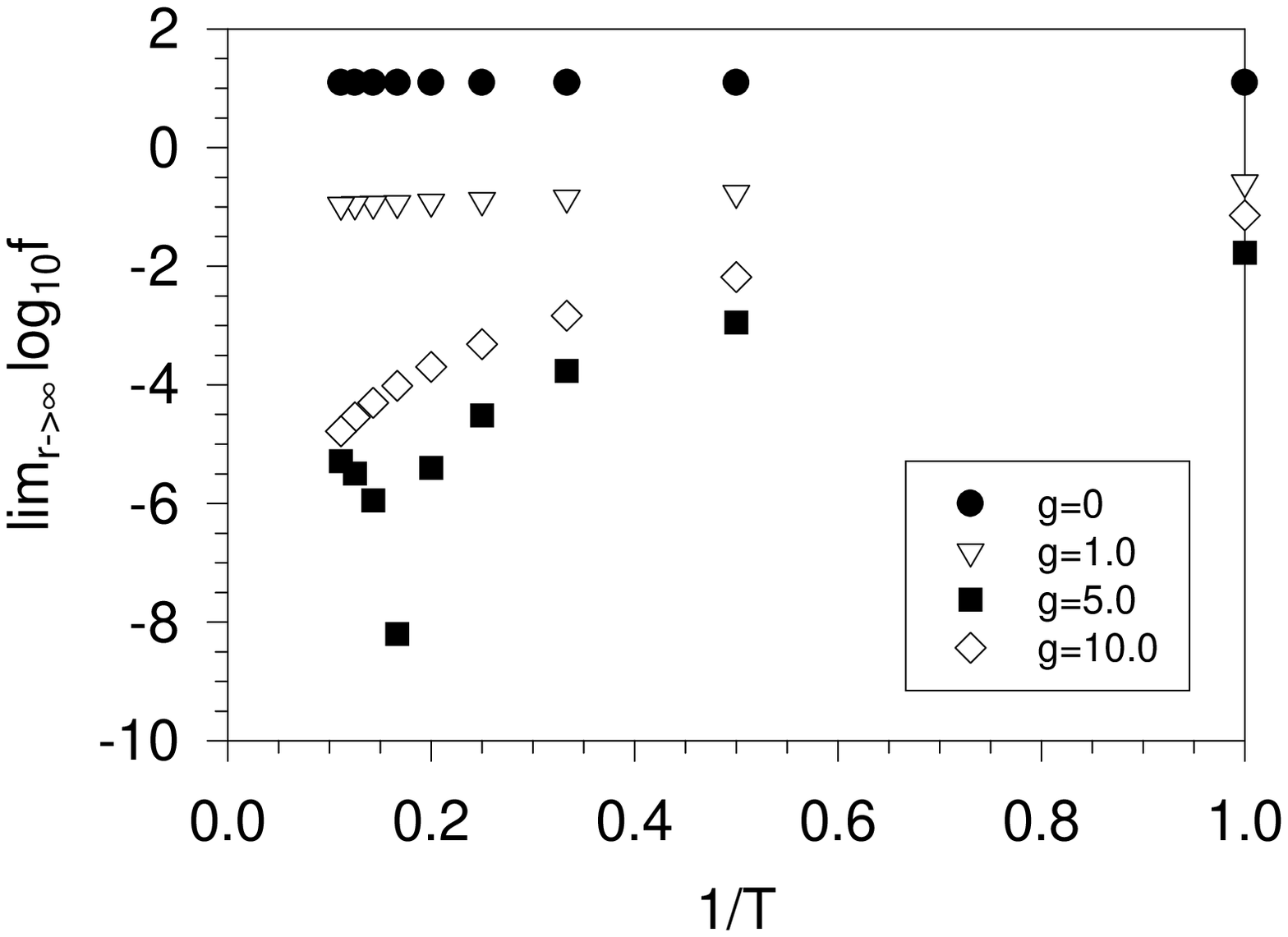,width=6cm}
\psfig{file=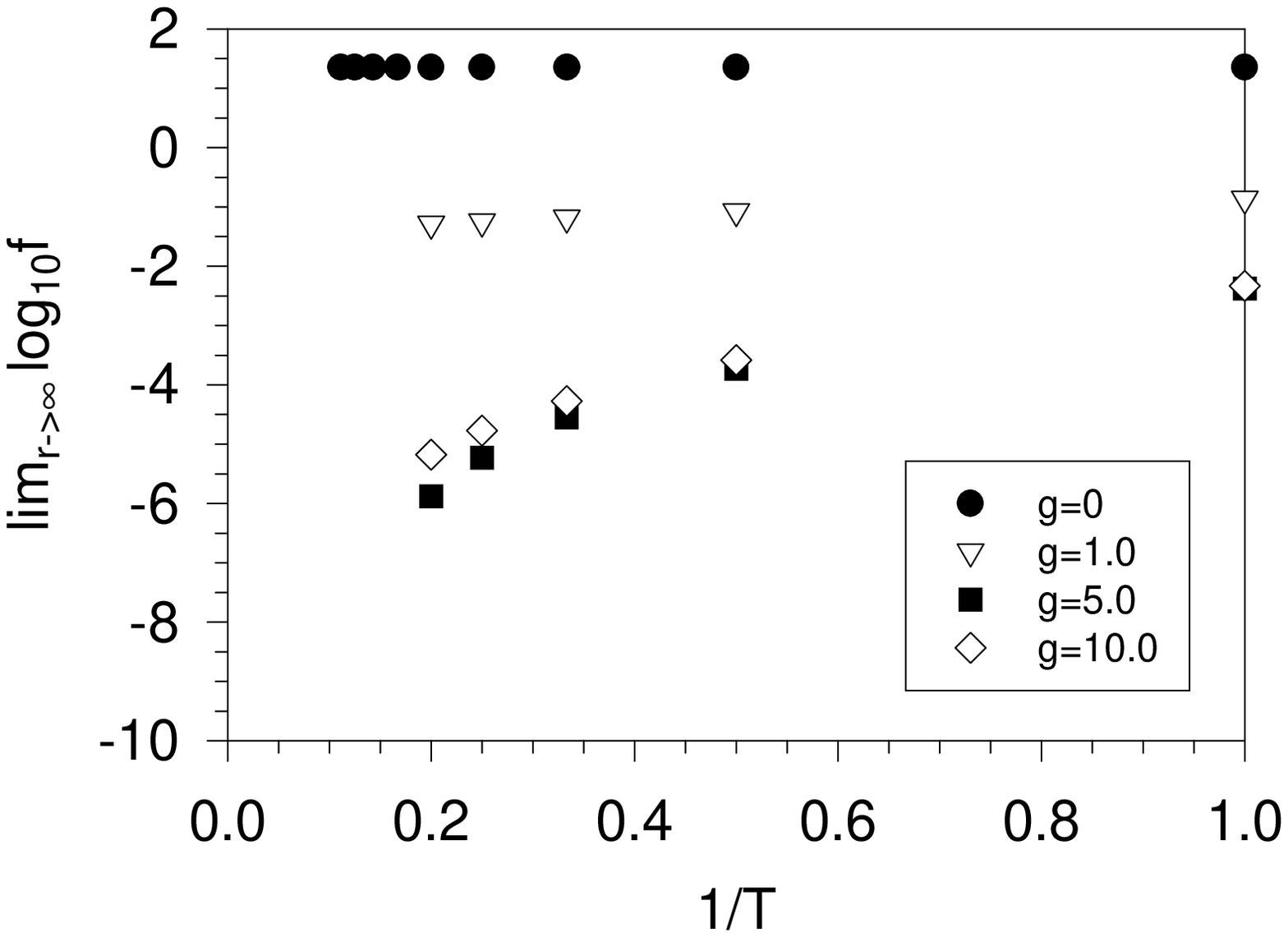,width=6cm} }
\caption{The large-$r$ limit of the log of
the correlation function $f\equiv r^5\langle T^{++}(x) T^{++}(0) \rangle
\left({x^- \over x^+} \right)^2 \frac{16\pi^3}{105}{{K^3 l} \over
\sqrt{-i}}$
vs.\ $1/T$ for [left](a) $K=5$ and [right](b) 
$K=6$ and for various values of the
coupling $g=g_{\rm YM} \sqrt{N_c l}/2\pi^{3/2}$.
\label{k5+6large}}
\end{figure}

It remains a challenge to rigorously test the conjectured
string/field theory correspondences. Although the so-called Maldacena
conjecture maybe the most exciting one, because it promises insight
into full four-dimensional Yang-Mills theories in the strong coupling
regime,
there are other interesting scenarios. For instance, it was conjectured that
the supergravity solutions corresponding to $p+1$ SYM theories are black
$p$-brane solutions, see {\em e.g.} Ref.~\cite{Itzhaki}.
Consequently, there are interesting testing scenarios also in
three-dimensional spacetime. Numerically, of course, things get more
difficult as the number of dimensions is increased.
On the way to the full four-dimensional problem, it may be worthwhile to
present 
our latest results on correlation functions in three dimensions; see
also \cite{now}. Fig.~\ref{k5+6large}(b) shows the correlator for
${\cal N}=1$ SYM(2+1) as a function of the distance $r$: it is
converging well with the {\em transverse} cut-off $T$.
To put things in perspective, we note that the construction of
the largest Hamiltonian matrix involved in this calculation requires a
Fock basis of approximately two million states. This is by a factor 100 more
than we used in the test of the Maldacena conjecture described in this
article, which itself was already substantially
better than the first feasibility study \cite{Anton98b}.

The correlator of the energy momentum
operator has been studied in conformal field theory in 2+1 dimensions
\cite{osp93}, and this provides a reference point for our results.
The structure of the correlators in conformal field theory is particularly
simple in the collinear limit $x_\perp\rightarrow 0$, and we therefore find
it convenient to work in this limit.
>From results in conformal field theory we expect that
correlators behave as $1/r^6$ at small $r$, where we are probing deep inside
the bound states.  We have confirmed this
$1/r^6$ behavior by an analytic calculation of the free-particle correlator
in the DLCQ formalism \cite{BPP}.

The contributions of individual bound states have a characteristic length
scale
corresponding to the size of the bound states.  On dimensional grounds one
can
show that the power behavior of the correlators are reduced by one power of
$r$; so
for individual bound states the correlator behaves like $1/r^5$ for small
$r$. It
then becomes a nontrivial check to see that at small $r$ the contributions
of the
bound states add up to give the expected $1/r^6$ behavior.  We find this
expected
result as well as the characteristic rapid convergence of SDLCQ at both
small and
intermediate values of $r$.

At large $r$ the
correlator is controlled by the massless states of the theory. In this
theory
there are two types of massless states. At zero coupling all the states of
the $1+1$
dimensional theory are massless, and for non-vanishing coupling
the massless states of the
$1+1$ theory are promoted to massless states of the $2+1$ dimensional theory
\cite{alp99b}. These states are BPS states and are exactly annihilated by
one of
the supercharges. This is perhaps the most interesting part of this
calculation
because the BPS masses are protected by the exact
supersymmetry of the numerical approximation and remain exactly zero at all
couplings. Commonly in modern field theory one uses the BPS states to
extrapolate
from weak coupling to strong coupling. While the masses of BPS states remain
constant as functions of the coupling, their wave functions certainly do
not. The
calculation of the correlator at large $r$ provides a window to the coupling
dependence of the BPS wave functions.  We find, however, that there is a
critical
coupling  where the correlator goes to zero, which depends on
the transverse resolution. A detailed study of this
critical coupling shows that it goes to infinity linearly with the square
root of the
transverse resolution. Below the critical coupling the correlator converges
rapidly
at large $r$. One possible explanation
is that this singular behavior signals the breakdown of
the SDLCQ calculation for the BPS wave function at couplings larger than the
critical
coupling. If this is correct, calculation of the BPS wave function at
stronger
couplings  requires higher transverse resolutions. We note that above the
critical
coupling (see Fig.~\ref{k5+6large} below)  we do find convergence of the
correlator
at large $r$ but at a significantly slower rate.


Let us now return to the details of the calculation.
We would like to compute a general expression of the form
\begin{equation}
F(x^+,x^-,x^\perp) = \langle 0| T^{++}(x^+,x^-,x^\perp)
T^{++}(0,0,0)|0 \rangle.
\end{equation}
Here we will calculate the correlator in the collinear limit, that is,
where $x^\perp = 0$. We know from conformal field theory
\cite{osp93}
calculations that this
will produce a much simpler structure.

The calculation is done by inserting a
complete set of intermediate states $| \alpha \rangle$,
\begin{eqnarray} \label{eq:F}
&&F(x^+,x^-,x^\perp=0) =
\\&&
 \sum_\alpha \langle 0| T^{++}(x^-,0,x^\perp=0)
| \alpha \rangle e^{-iP^-_\alpha x^+} \langle \alpha | T^{++}(0,0,0)
|0 \rangle. \nonumber
\end{eqnarray}
with energy eigenvalues $P^-_\alpha$.
The momentum operator $T^{++}(x)$ is given by
\begin{equation}
T^{++}(x) =  {\rm tr} \left[ (\partial_- \phi)^2 + {1 \over 2} \left(i
\psi \partial_- \psi  - i  (\partial_- \psi) \psi
\right)\right]=T^{++}_B(x)+T^{++}_F(x)\,.
\end{equation}
In terms of the mode operators, we find
\begin{equation}
T^{++}(x^+,x^-,0) | 0 \rangle = {1 \over 2L l} \sum_{n,m}\,\,
\sum_{n_\perp,m_\perp}
T(n,m) e^{-i(P_n^+ + P_m^+)x^-}
   |0\rangle\,,
\end{equation}
where the boson and fermion contributions are given by
\begin{equation}
\frac{L}{\pi}T^{++}_B(n,m) | 0 \rangle = {\sqrt{n m} \over 2}
    {\rm tr} \left[ a^\dagger_{ij}(n,n_\perp) a^\dagger_{ji} (m,m_\perp)
   \right]| 0\rangle
\end{equation}
and
\begin{equation}
\frac{L}{\pi}T^{++}_F(n,m) | 0 \rangle = {(n-m) \over 4}
{\rm tr}\left[b^\dagger_{ij}(n,n_\perp) b^\dagger_{ji}(m,m_\perp)
\right] | 0 \rangle\,.
\end{equation}
Given each $|\alpha\rangle$, the matrix elements in (\ref{eq:F}) can
then be evaluated, and the sum computed.

First, however, it is instructive to do the calculation where the states
$|\alpha\rangle$
are a set of free particles with mass $m$. The boson contribution is
\begin{eqnarray}
&&F(x^+,x^-,0)_B= \sum_{n,m,s,t} \left(\frac{\pi}{4L^2 l}\right)^2
e^{-iP^-_n x^+ -iP^+_n x^--iP^-_m x^+ -iP^+_m x^-}\\
&&\times \sqrt{mnst} \, \langle 0|{\rm tr}[a(n,n_\perp) a(m,m_\perp)] {\rm
tr}[a^\dagger(s,s_\perp)
a^\dagger(t,t_\perp)] |0\rangle,\nonumber
\end{eqnarray}
where the sum over $n$ implies sums over both $n$ and $n_\perp$, and
\begin{equation}
P^-_n=\frac{m^2 +\left(2 n_\perp \pi/l \right)^2}{2 n \pi/L}
\quad  {\rm and} \quad P_n^+=\frac{n \pi}{L}\,.
\end{equation}
The sums can be converted to integrals which can be explicitly evaluated,
and we
find
\begin{equation}
F(x^+,x^-,0)_B=\frac{i}{2(2\pi)^3} m^5 \left(\frac{x^+}{x^-} \right)^2
\frac{1}{x} K^2_{5/2}(mx)\,,
\end{equation}
where $x^2 =2x^-x^+$.
Similarly for the fermions we find
\begin{eqnarray}
&&F(x^+,x^-,0)_F= \sum_{n,m,s,t} \left(\frac{\pi}{8L^2 l}\right)^2
 e^{-iP^-_n x^+ -iP^+_n x^--iP^-_m x^+ -iP^+_m x^-}\,\\ \nonumber
&&\times (m-n) (s-t) \,\langle 0|{\rm tr}[b(n,n_\perp) b(m,m_\perp)] {\rm
tr}[b^\dagger(s,s_\perp)
b^\dagger(t,t_\perp)] |0\rangle.
\end{eqnarray}
After doing the integrals we obtain
\begin{equation}
F(x^+,x^-,0)_F=\frac{i}{4(2\pi)^3} \frac{m^5}{x} 
\left(\frac{x^+}{x^-} \right)^2
\left[ K_{7/2}(mx)K_{3/2}(mx)-K^2_{5/2}(mx) \right]\,.
\end{equation}
We can continue to Euclidean
space by taking $r = \sqrt{2 x^+ x^-}$ to be real, and, finally, in the
small-$r$ limit we find
\begin{equation}
\left(\frac{x^-}{x^+} \right)^2 F(x^+,x^-,0)=\frac{-3 i}{8 (2\pi)^2}
\frac{1}{r^6}\,,
\end{equation}
which exhibits the expected $1/r^6$ behavior.


Now  let us return to the calculation using the bound-state solution
obtained from SDLCQ. It is convenient to write
\begin{equation} \label{eq:Fdiscrete}
F(x^+,x^-,0)=\sum_{n,m,s,t} \,\,\left(\frac{\pi}{2L^2 l}\right)^2
\langle 0|\frac{L}{\pi}T(n,m) e^{-iP^-_{op}x^+-iP^+x^-}
            \frac{L}{\pi} T(s,t)| 0 \rangle\,,
\end{equation}
where $P^-_{op}$ is the Hamiltonian operator.
We again insert a complete set of bound states $|\alpha\rangle $
with light-cone energies $P_\alpha^-=(M_\alpha^2+P_\perp^2)/P^+$
at resolution K (and therefore $P^+=\pi K/L$) and with total
transverse momentum $P_\perp=2N_\perp\pi/l$. We also define
\begin{equation}
|u\rangle = N_u \frac{L}{\pi}
  \sum_{n,m}\delta_{n+m,K}\delta_{n_\perp+m_\perp,N_\perp}T(n,m) |0
\rangle\,,
\end{equation}
where $N_u$ is a normalization factor such that $\langle u|u \rangle=1$.
It is straightforward to calculate the normalization, and we find
\begin{equation}
\frac{1}{N_u^2}=\frac{K^3}{8} (1-\frac{1}{K}) (2T+1)\,.
\end{equation}
The correlator (\ref{eq:Fdiscrete}) becomes
\begin{equation}
F(x^+,x^-,0)=\sum_{K,N_\perp,\alpha} \left(\frac{\pi}{2L^2 l}\right)^2
e^{-iP^-_\alpha
x^+-iP^+x^-} \frac{1}{N_u^2} |\langle u|\alpha\rangle |^2\,.
\end{equation}

We will calculate the matrix element $\langle u|\alpha \rangle$ at fixed
longitudinal resolution $K$ and transverse momentum $N_\perp=0$.
Because of transverse boost invariance
the matrix element does not contain any explicit dependence on $N_\perp$.
To leading order in
$1/K$ the explicit dependence of the matrix element on $K$ is $K^3$;
it also contains a factor of $l$, the transverse length scale.
To separate these dependencies, we write $F$ as
\begin{equation}
F(x^+,x^-,0)=\frac{1}{2 \pi} \sum_{K,N_\perp,\alpha} \frac{1}{2L}
\frac{1}{l} \left(\frac{\pi K}{L}\right)^3
  e^{-iP^-_\alpha x^+ -iP^+ x^-}
  \frac{|\langle u|\alpha\rangle|^2}{l K^3 |N_u|^2} \,.
\end{equation}

We can now do the sums over $K$ and $N_\perp$ as integrals over the
longitudinal and transverse momentum components
$P^+=\pi K/L$ and $P^\perp= 2 \pi N_\perp/l$.  We obtain
\begin{equation}
\frac{1}{\sqrt{-i} }\left(\frac{x^-}{x^+} \right)^2 F(x^+,x^-,0)
=\sum_\alpha 
\frac{1}{2 (2\pi)^{5/2}}\frac{M_\alpha^{9/2}}{\sqrt{r}}K_{9/2}(M_\alpha r)
\frac{|\langle u|\alpha\rangle|^2}{lK^3 |N_u|^2}.
\label{master2}
\end{equation}
In practice, the full sum over $\alpha$ is approximated by a Lanczos
\cite{Lanczos} iteration technique \cite{this work,Haydock} that eliminates
the need for full diagonalization of the Hamiltonian matrix.
For the present case, the number of iterations required was on the
order of 1000.

Looking back at the calculation for the free particle, we see that there are
two independent sums over transverse momentum, after the contractions are
performed. One would expect that the transverse dimension is
controlled by the dimensional scale of the bound state ($R_B$) and therefore
the correlation should scale like $1/ r^4 R_B^2$. However, because of
transverse boost invariance, the matrix element must be independent of the
difference of the transverse momenta and therefore must scale as
$1/r^5R_B$.

There are three commuting
$Z_2$ symmetries.  One of them is the parity in the transverse direction,
\begin{equation}\label{parity}
P: a_{ij}(k,n^\perp)\rightarrow -a_{ij}(k,-n^\perp),\qquad
      b_{ij}(k,n^\perp)\rightarrow b_{ij}(k,-n^\perp).
\end{equation}
The second symmetry \cite{Kutasov93} is with respect to the operation
\begin{equation}\label{Z2}
S: a_{ij}(k,n^\perp)\rightarrow -a_{ji}(k,n^\perp),\qquad
      b_{ij}(k,n^\perp)\rightarrow -b_{ji}(k,n^\perp).
\end{equation}
Since $P$ and $S$ commute with each other, we need only one additional
symmetry $R=PS$ to close the group.
Since $Q^-$, $P$ and $S$ commute with each other, we can diagonalize them
simultaneously. This allows us to diagonalize the supercharge separately  in
the sectors with fixed $P$ and $S$ parities and thus reduce the size
of matrices. Doing this one finds that the roles of $P$ and $S$ are
different.
While all the eigenvalues are usually broken into non-overlapping $S$-odd
and
$S$-even sectors \cite{bdk93}, the $P$ symmetry leads to a double
degeneracy of massive states (in addition to the usual boson-fermion
degeneracy
due to supersymmetry).

%
\begin{figure}
\centerline{
\psfig{file=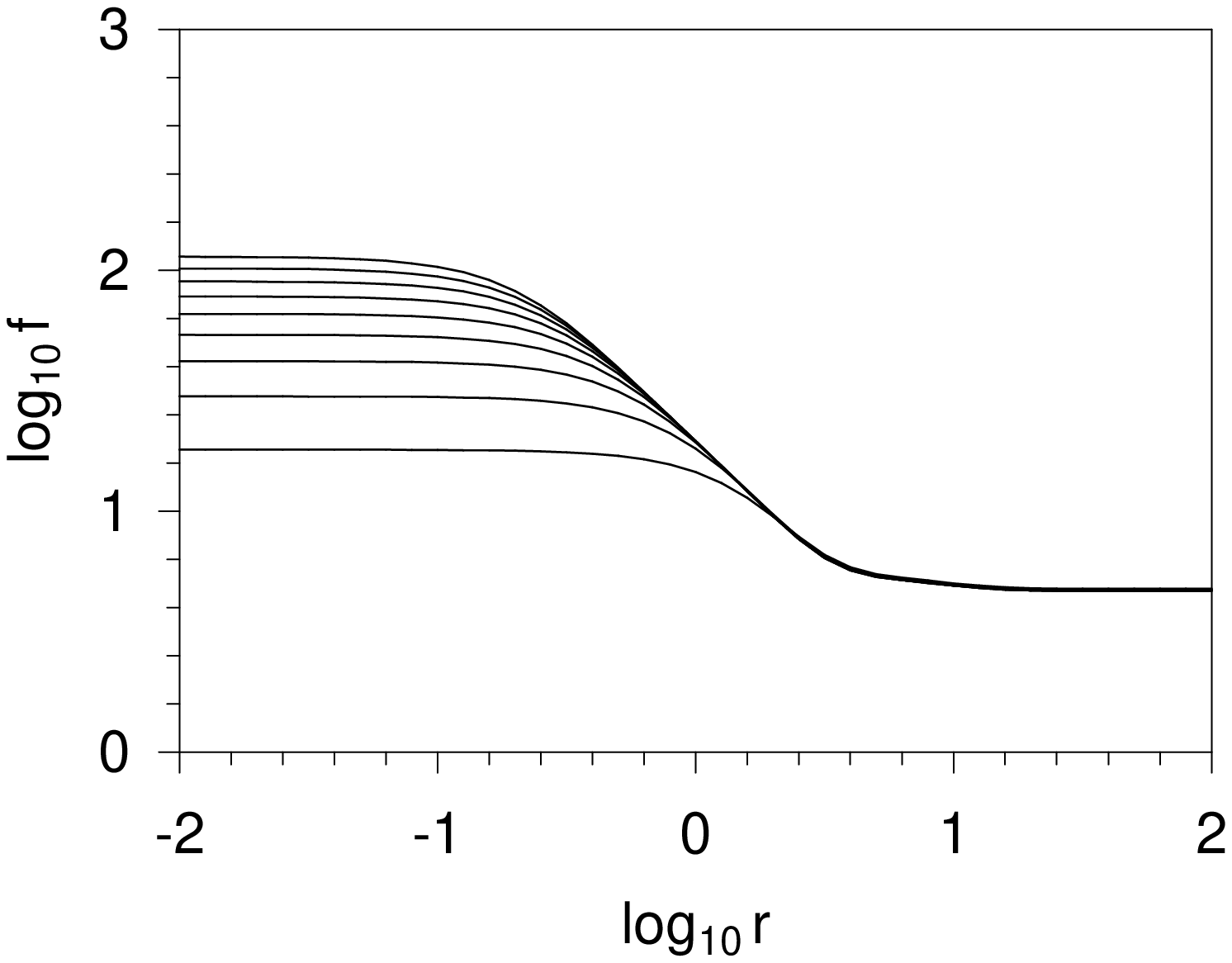,width=6cm}
\psfig{file=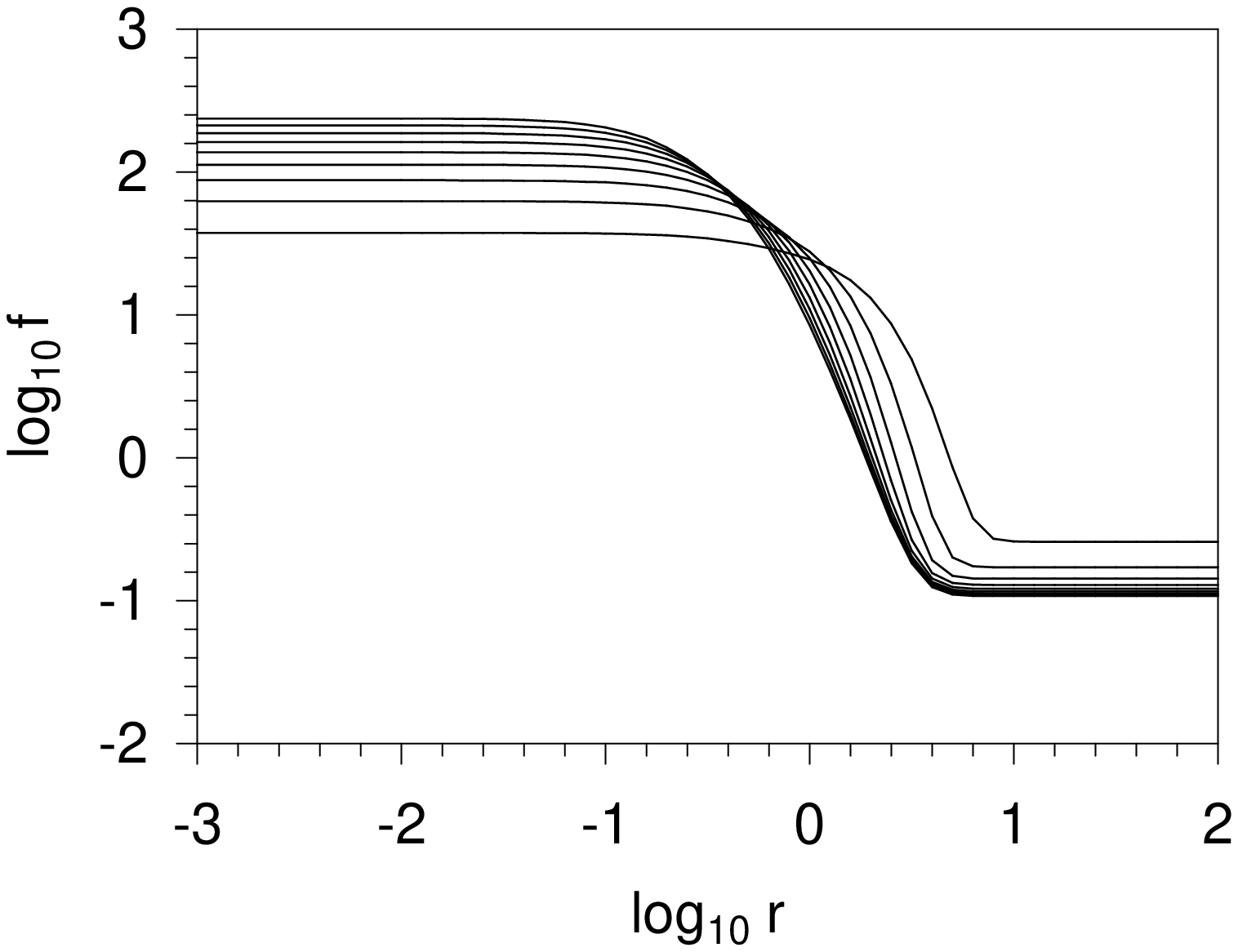,width=6cm}}
\caption{The log-log
plot of the correlation function $f\equiv r^5\langle T^{++}(x) T^{++}(0)
\rangle
\left({x^- \over x^+} \right)^2 \frac{16\pi^3}{105}{{K^3 l} \over
\sqrt{-i}}$
vs.\ $r$. 
Left: (a) in units where $g=g_{\rm YM} \sqrt{N_c l}/2\pi^{3/2} = 0.10$
for $K=4$
and
$T=1$ to 9; Right: 
(b) in units where $g=g_{\rm YM} \sqrt{N_c l}/2\pi^{3/2} = 1$ for $K=5$
and
$T=1$ to 9.
\label{smallr}}
\end{figure}
%

{\bf Numerical Results.}
The first important numerical test is the small-$r$ behavior of the
correlator.
Physically we expect that at small $r$ the bound states should behave as
free
particles, and therefore the correlator should have the behavior of the free
particle correlator which goes like $1/r^6$. We see in (\ref{master2}) that
the
contributions of each of the bound states behaves like $1/r^5$.
Therefore, to get the $1/r^6$ behavior of the free theory, the bound
states must work in concert at small $r$. It
is clear that this cannot work all the way down to $r=0$
in the numerical calculation. At very small $r$
the most massive state allowed by the numerical approximation will
dominate, and the correlator must behave like
$1/r^5$. To see what happens at slightly larger $r$ it is useful to
consider the behavior at small coupling. There, the larger masses go
like
\begin{equation}
M_\alpha \simeq \sum_i \frac{(k^\perp_i)^2}{2P^+}\,.
\end{equation}
Consequently, as we remove the $k^\perp$ cutoff, {\em i.e.}~increase the
transverse
resolution $T$, more and more massive bound states will contribute, and the
dominant one will take over at smaller and smaller $r$ leading to the
expected $1/r^6$. This is exactly what we see happening in
Fig.~\ref{smallr} at weak coupling with longitudinal resolution $K=4$ and 5.

The correlator converges from below at small $r$ with increasing $T$, and
in the region $ -0.5 \leq \log r \leq 0.5 $ the plot of $r^5$ times the
correlator falls like $ 1/r$. In Fig.~\ref{k5+6large}
at resolution $K=5$ we see
the same behavior for strong coupling ($g =g_{\rm YM} \sqrt{N_c
l}/2\pi^{3/2} =10$)
but now at smaller $r$ ($\log r \simeq -0.5 $) as one would expect.
%
\begin{figure}
\centerline{
\psfig{file=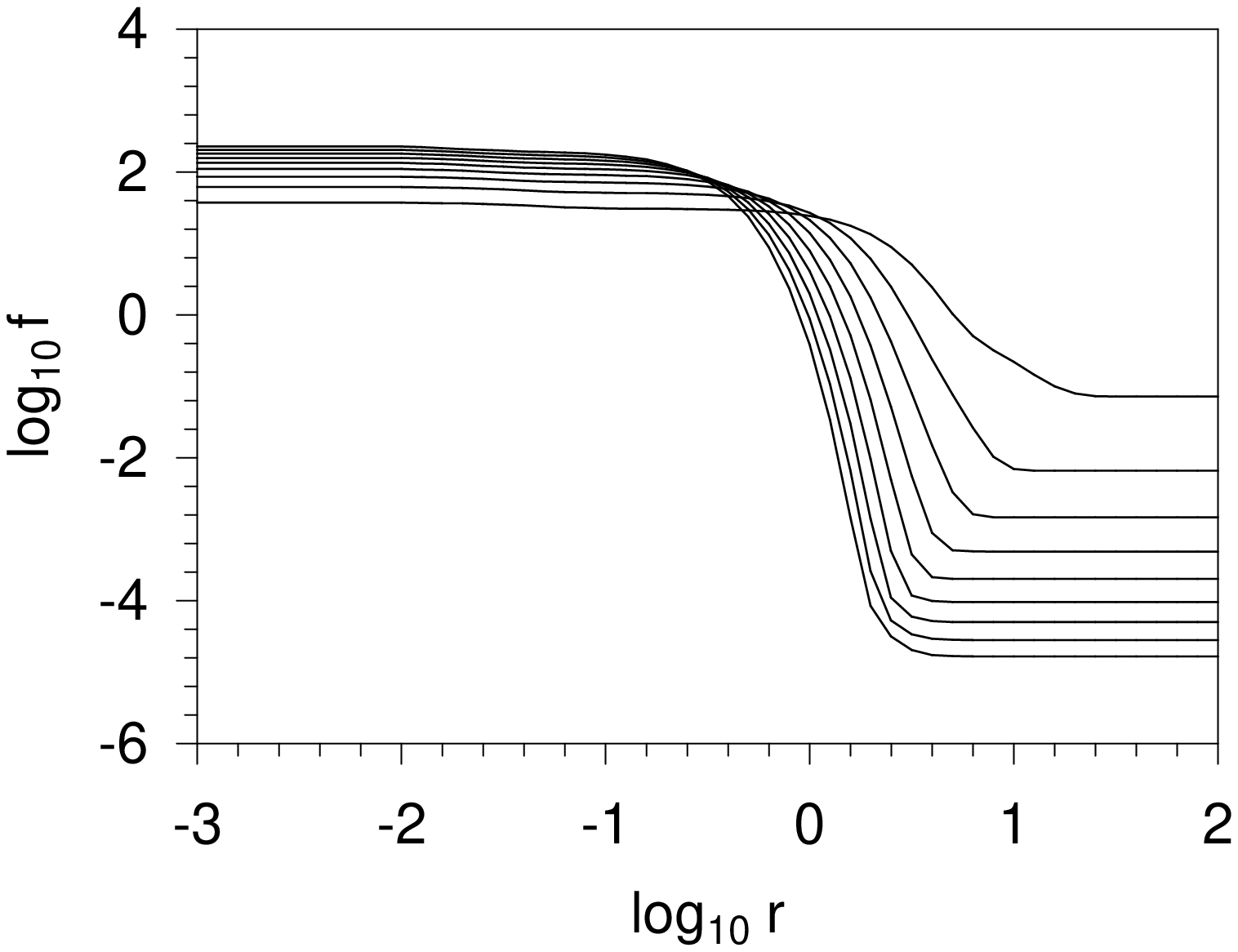,width=6cm}
\psfig{file=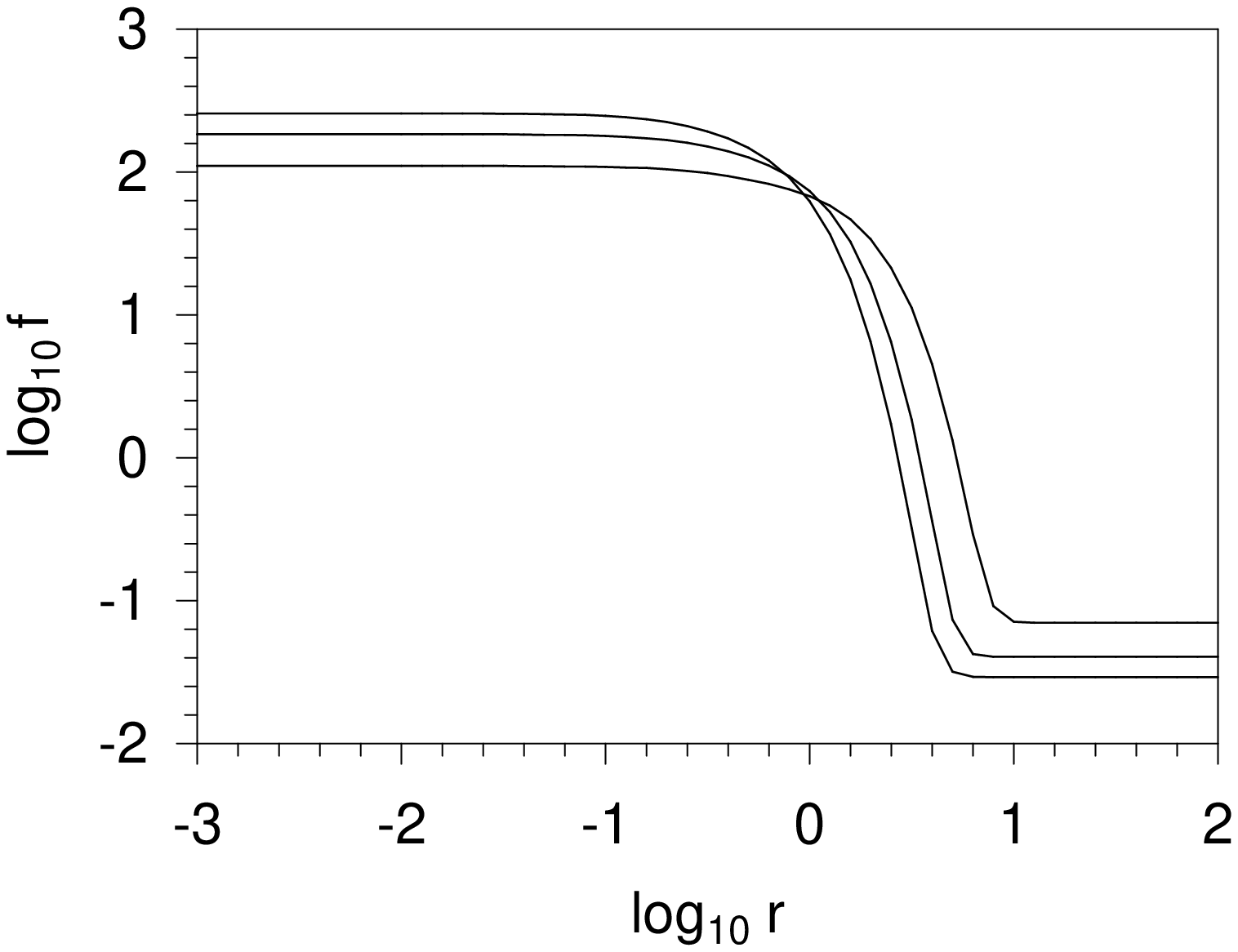,width=6cm}}
\caption{Left: (a) Same as Fig.~\ref{smallr}(b), but for
$g=g_{\rm YM} \sqrt{N_c l}/2\pi^{3/2}= 10$.
Right: (b) Log-Log
plot of the {\em three-dimensional} correlation function $f\equiv r^5
\langle T^{++}(x) T^{++}(0) \rangle
\left({x^- \over x^+} \right)^2 {16\pi^3 K^3 l\over 105\sqrt{-i}}$
vs. $r$ for $g=g_{YM}^2 \sqrt{N_c l} /2\pi^{3/2} = 1.0$ for $K=6$ and
$T=1$ to $5$.}
\vspace*{-0.5cm}
\end{figure}
Again at strong
coupling we see that the correlator converges quickly and from
below in $T$.  All indications are that at small $r$ the correlators are
well
approximated by SDLCQ, converge rapidly, and show the behavior that one
would expect on general physical grounds. This gives us confidence to go
on to investigate the behavior at large $r$.

The behavior for large $r$ is governed by the massless states. From earlier
work \cite{hhlp99,alp99b} on the spectrum of this theory we know that
there are two types of massless states.  At $g=0$ the
massless states are  a reflection of all the states of the dimensionally
reduced theory in $1+1$. In 2+1 dimensions these states behave as $g^2
M^2_{1+1}$. We expect therefore that for $g \simeq 0$ there should be no
dependence of the correlator on the transverse momentum cutoff
$T$ at large $r$. In Fig.~\ref{smallr}(a) this behavior is clearly evident.

At all couplings there are exactly massless states which are the BPS states
of
this theory, which has zero central charge. These states are destroyed by
one
supercharge, $Q^-$, and not the other, $Q^+$.  From earlier
work \cite{hhlp99} on the spectrum we saw that the number of BPS states is
independent of the transverse resolution and equal to $2K-1$. Since these
states are exactly massless at all resolutions, transverse and longitudinal
convergence of these states cannot be investigated using the spectrum.
These states do have a complicated dependence on the coupling $g$ through
their wave function, however. This is a
feature so far not encountered in DLCQ \cite{BPP}.
In previous DLCQ calculations one always
looked to the convergence of the spectrum as a measure of the convergence of
the
numerical calculation. Here we see that it is the correlator at large $r$
that provides a window to study the convergence of the  wave functions of
the BPS
states. In Fig.~\ref{k5+6large} we see that the correlator converges
from above at large $r$ as we increase $T$.

We also note that the correlator at large $r$ is significantly smaller than
at
small $r$, particularly at strong coupling. In our initial study of the BPS
states
\cite{alp99b} we found that at strong coupling the average number of
particles
in these BPS states is large. Therefore the two particle
components, which are the only components the $T^{++}$ correlator sees, are
small.

\begin{figure}
\centerline{
\psfig{file=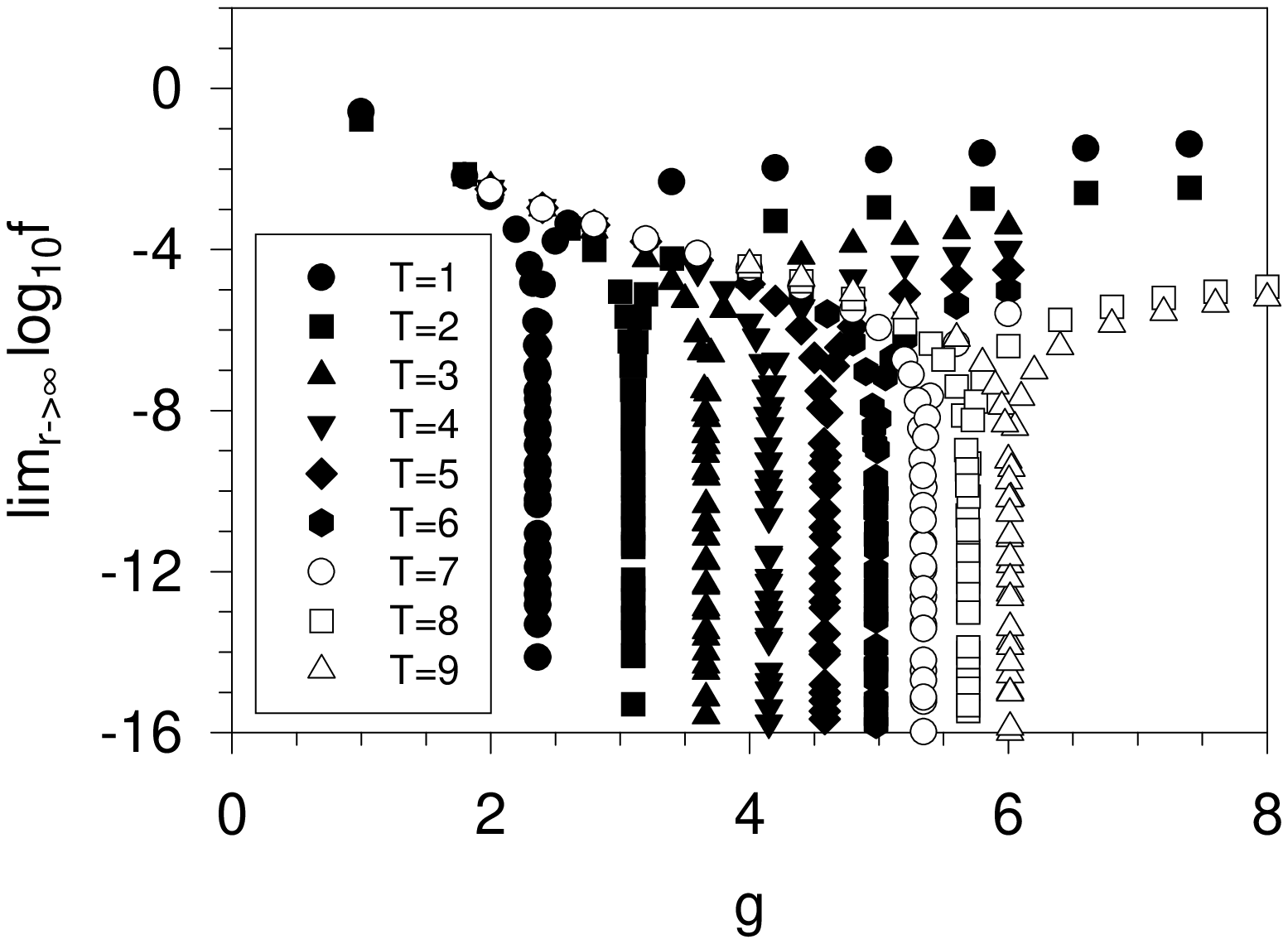,width=6cm}
\psfig{file=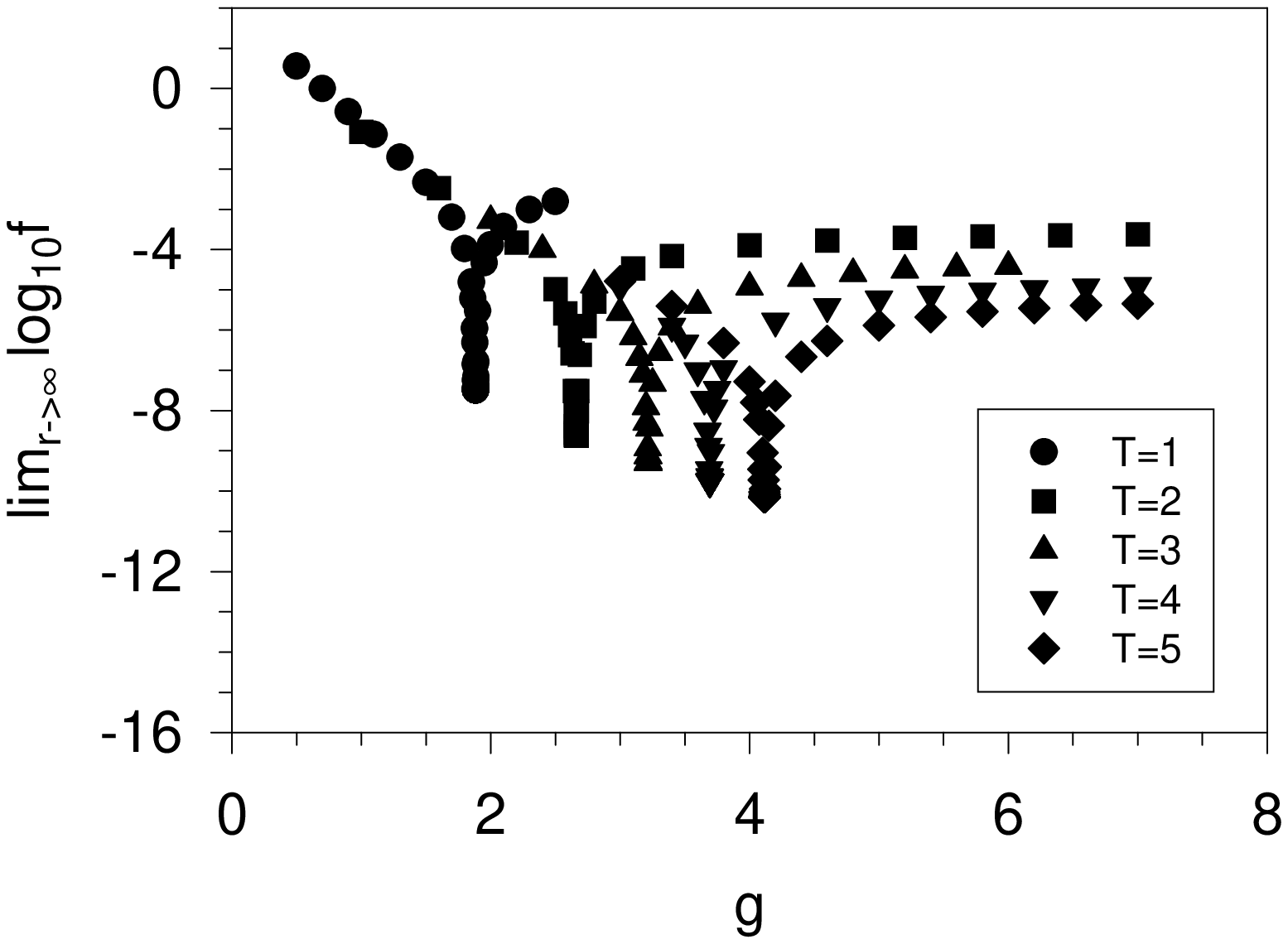,width=6cm}}
\caption{The large-$r$ limit of
the correlation function $f\equiv r^5\langle T^{++}(x) T^{++}(0) \rangle
\left({x^- \over x^+} \right)^2 \frac{16\pi^3}{105}{{K^3 l} \over
\sqrt{-i}}$ 
vs.\ $g=g_{\rm YM} \sqrt{N_c l}/2\pi^{3/2}$ for [left](a) $K=5$ [right](b) 
$K=6$ and
for various values of the transverse resolution $T$.\label{k5+6crit}}
\end{figure}

The coupling dependence of the large-$r$ limit of the correlator is much
more 
interesting than we would have expected based on our previous work on the
spectrum. To see this
behavior we study the large-$r$ behavior of the correlator at fixed $g$ as a
function of the transverse resolution $T$ and at fixed $T$ as a function of
the
coupling $g$. We see a hint that something unusual is occurring in
Fig.~\ref{k5+6large}. For values of the coupling up to about $g=1$ we see
the
typical rapid convergence in the transverse momentum cutoff; however,
at larger coupling
the convergence appears to deteriorate, and we see that for
$g=5$ the correlator is smaller than at $g=10$. We see this same behavior
at both $K=5$ and $K=6$. We do not see this behavior at
$K=4$, but it is not unusual for effects to appear only at a large enough
resolution in SDLCQ.
In Fig.~\ref{k5+6crit} we see that the correlator does not in fact
decrease monotonically with $g$ but rather has a singularity at a
particular value of the coupling which is a function of $K$ and $T$.
Beyond the singularity the correlator again appears to behave well.

\begin{figure}
\begin{tabular}{cc}
\psfig{file=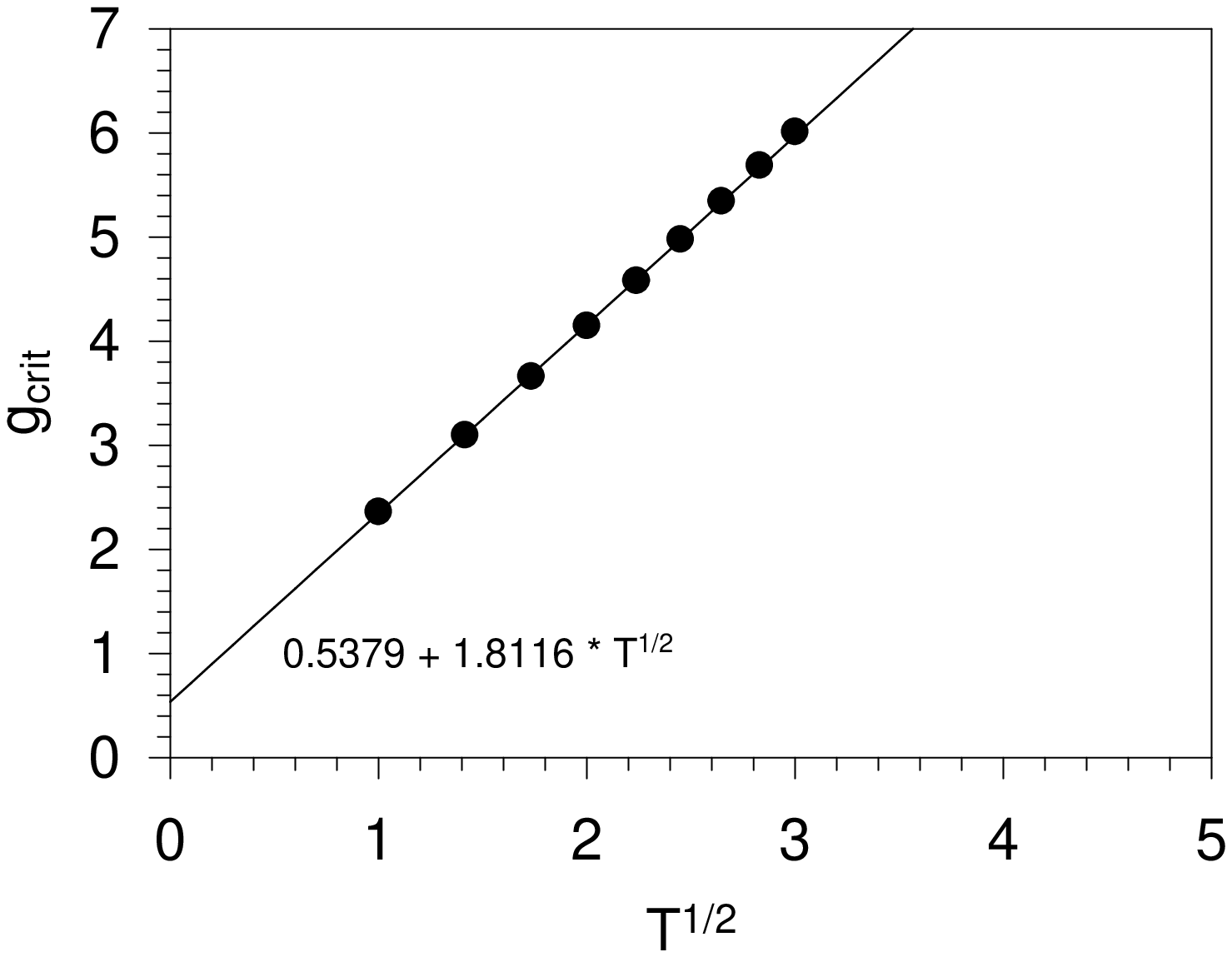,width=5.8cm}  &
\psfig{file=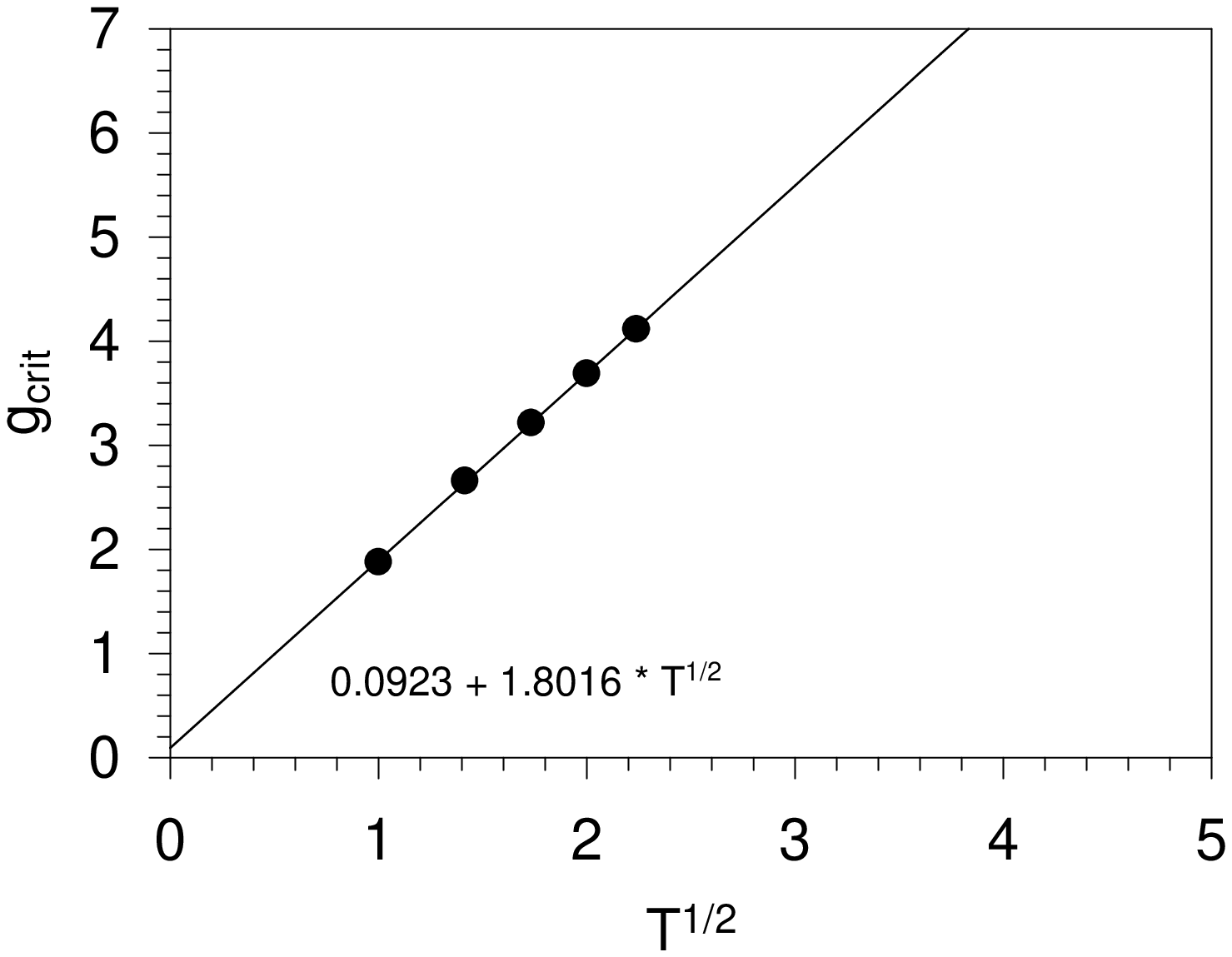,width=5.8cm}  \\
(a) & (b)
\end{tabular}
\caption{
Critical coupling $g_{\rm crit}$ versus $\sqrt{T}$ for (a) $K=5$ and (b)
$K=6$.
\label{k5+6critfit2}}
\end{figure}

If we plot the `critical' couplings, at which the correlator goes to zero,
versus $\sqrt{T}$, as in Fig.~\ref{k5+6critfit2},
we see that they lie on a straight
line, {\em i.e.}~this coupling is a linear function
of $\sqrt{T}$ in both cases, $K=5$ and 6.
Consequently, the `critical' coupling goes to infinity in the
transverse continuum limit.
It appears as though we have encountered a finite transverse cutoff effect.
The most
likely conclusion is that our numerical calculation of the BPS wave function
is only
valid for $g < g_{\rm crit}(T)$. While the large-$r$ correlator does
converge above
the critical coupling, it is unclear at this time if it has any
significance. It
might have been expected that one would need larger and larger transverse
resolution
to probe the strong coupling region, the occurrence of the singular behavior
that we
see is a surprise, and we have no detailed explanation for it at this time.
We see
no evidence of a singular behavior at small or intermediate $r$. This
indicates, but
does not prove, that our calculations of the massive bound states is valid
at all
$g$.

We do not seem to see a region dominated by the massive bound states, that
is,
a
region where $r$ is large enough that we see the structure of the
bound states but small enough that the correlator is not dominated by the
massless
states of the theory. Such a region might give us other important
information
about this theory.

\section{Conclusions}

In this note we reported on progress in an attempt to rigorously test
the conjectured equivalence of two-dimensional ${\cal N}=(8,8)$
supersymmetric Yang-Mills theory and a
system of $D1$-branes in string theory.
Within a well-defined non-perturbative calculation,
we obtained results that are within 10-15\% of results
expected from the Maldacena conjecture.
The results are still not conclusive, but they definitely
point in right direction.
Compared to previous work \cite{Anton98b}, we included orders of magnitude
more states in our calculation and
thus greatly improved the testing conditions.
We remark that improvements of the code and the numerical method
are possible and under way.
During the calculation we noticed that contributions
to the correlator come from only a small number of terms.
An analytic understanding of this phenomenon
would greatly accelerate calculations.
We point out that in principle we
could study the proper $1/r$ behavior at large $r$ by
computing $1/N_c$ corrections, but this interesting calculation
would mean a huge numerical effort.

In this work we also discussed the calculation of the stress-tensor 
correlator 
$\langle 0|T^{++}(x) T^{++}(0)|0\rangle$ in ${\cal N}=1$ SYM in 2+1
dimension at large
$N_c$ in the collinear limit.  We find that the free-particle correlator
behaves
like $1/r^6$, in agreement with results from conformal field theory.
The contribution from an individual bound state is found to
behave like $1/r^5$, and at small $r$ such contributions conspire to
reproduce
the conformal field theory result $1/r^6$. We do not seem to
find an intermediate region in
$r$ where the correlator behaves as $1/r^5$, reflecting the behavior of the
individual massive bound states.

At large
$r$ the correlator is dominated by the massless BPS states of the theory. We
find
that as a function of $g$ the large-$r$ correlator has a critical value
of $g$ where it abruptly drops to zero.
We have investigated this singular behavior and find that at fixed
longitudinal
resolution the critical coupling grows linearly with $\sqrt{T}$. We
conjecture that
this critical coupling signals the breakdown of SDLCQ at sufficiently strong
coupling
at fixed transverse resolution, $T$. While this might not be surprising in
general,
it is surprising that the behavior appears in the BPS wave functions and
that we see
no sign of this behavior in the massive states. We find that above the
critical
coupling the correlator still converges but significantly slower. It is
unclear at
this time if we should attach any significance to the correlator in this
region. 

This calculation emphasizes the importance of BPS wave functions which carry
important
coupling dependence, even though the mass eigenvalues are independent of the
coupling. We will discuss the spectrum, the wavefunctions and associated
properties of all the low energy bound states of
${\cal N}=1$ SYM  in 2+1 dimensions in a subsequent paper \cite{hpt01}.

A number of computational improvements have been implemented in our code
to allow
us to make these detailed calculations. The code now fully utilizes the
three known
discrete symmetries of the theory, namely supersymmetry, transverse parity
$P$, 
Eq.~(\ref{parity}), and the $Z_2$ symmetry $S$,
Eq.~(\ref{Z2}).
This reduces the dimension of the Hamiltonian matrix by a factor of 8.
Other, more efficient storage techniques allow us
to handle on the order of 2,000,000 states in this calculation, which has
been
performed on a single processor Linux workstation. Our improved storage
techniques
should allow us to expand this calculation to include higher supersymmetries
without a significant expansion of the code or computational power. We
remain hopeful that porting to a parallel machine will allow us to tackle
problems
in full 3+1 dimensions.

\section*{Acknowledgments}

We would like to thank the organizers for the opportunity to speak
at this workshop. This work was supported in part by the US Department of
Energy.


\begin{chapthebibliography}{99}
%
\bibitem{Maldacena} J. Maldacena,
{\em Adv.\ Theor.\ Math.\ Phys.} {\bf 2} (1998) 231,
hep-th/9711200.
%
\bibitem{lup99}
O. Lunin and S. Pinsky,
in the proceedings of 11th International Light-Cone School and Workshop:
New Directions in Quantum Chromodynamics and 12th Nuclear Physics Summer
School and Symposium (NuSS 99), Seoul, Korea, 26 May - 26 Jun 1999
(New York, AIP, 1999), p.~140,  hep-th/9910222.

%
\bibitem{Itzhaki}
N. Itzhaki, J. Maldacena, J. Sonnenschein, and S. Yankielowicz,
{\em Phys.\ Rev.}\ {\bf D58} (1998) 046004, hep-th/9802042.
%
\bibitem{Anton98b} F. Antonuccio, O. Lunin, S. Pinsky, and A. Hashimoto,
JHEP {\bf 07} (1999) 029.
%
\bibitem{ItzhakiHashimoto} A. Hashimoto and N. Itzhaki,
{\em Phys.~Lett.~}{\bf B465} (1999) 142--147, hep-th/9903067.
%
\bibitem{Gubser} S. S. Gubser, I. R. Klebanov, and A. M. Polyakov,
{\em Phys.\ Lett.}\ {\bf B428} (1998) 105, hep-th/9802109.
%
\bibitem{Witten} E. Witten,
{\em Adv.\ Theor.\ Math.\ Phys.}\ {\bf 2} (1998) 253, hep-th/9802150.
%
\bibitem{KRvN}
H.~J. Kim, L.~J. Romans, and P.~van Nieuwenhuizen,
{\em Phys. Rev.} {\bf D32} (1985) 389--399.
%
\bibitem{krasnitz1}
M.~Krasnitz and I.~R. Klebanov,
{\em Phys. Rev.} {\bf D56} (1997) 2173--2179,
hep-th/9703216.
%
\bibitem{krasnitz2}
S.~S. Gubser, A.~Hashimoto, I.~R. Klebanov, M.~Krasnitz,
{\em Nucl. Phys.} {\bf B526} (1998) 393,
hep-th/9803023.
%
\bibitem{ferrara}
S.~Ferrara, C.~Fronsdal, and A.~Zaffaroni,
{\em Nucl. Phys.} {\bf B532} (1998) 153, hep-th/9802203.
%
\bibitem{Zamo} A.B. Zamolodchikov,
{\em JETP Lett.~}{\bf 43} (1986) 730;
 {\em Sov.J.Nucl.Phys.~}{\bf 46} (1987) 1090.
%
\bibitem{now} J.R. Hiller, S. Pinsky, U. Trittmann,
{\sl Two-Point Stress-Tensor Correlator in ${\cal N}=1$ SYM(2+1)},
hep-th/0101120.
%
\bibitem{Sakai95}
Y. Matsumura, N. Sakai, and T. Sakai,
{\em Phys.\ Rev.}\ {\bf D52} (1995) 2446,
hep-th/9504150.
%
\bibitem{hak95}
A. Hashimoto and I.R. Klebanov,
{\em Nucl.\ Phys.}\ {\bf B434} (1995) 264.
%
\bibitem{BPP} S.J. Brodsky, H.-C. Pauli, and S.S. Pinsky,
{\em Phys.\ Rept.}\ {\bf301} (1998) 299.
%
\bibitem{alp99b} F. Antonuccio, O. Lunin, and S. Pinsky,
{\em Phys.\ Rev.}\ {\bf D59} (1999) 085001.
%
\bibitem{alp98}
F.~Antonuccio, O.~Lunin, S.~Pinsky,
  {\em Phys.\ Lett.}\ {\bf B429} (1998) 327, hep-th/9803027.
%
\bibitem{adi1}
A.~Armoni, Y.~Frishman, and J.~Sonnenschein, 
{\em Phys. Lett.} {\bf B449} (1999) 76,
{{hep-th/9807022}}.

\bibitem{adi2}
A.~Armoni, Y.~Frishman, and J.~Sonnenschein, 
{{hep-th/9903153}}.
%
\bibitem{alp98a}
F.~Antonuccio, O.~Lunin, and S.~Pinsky,
{\em Phys.\ Rev.}\ {\bf D58} (1998) 085009.
%
\bibitem{dlcq22} F.~Antonuccio, H.-C.~Pauli,
S.~Pinsky, S.~Tsujimaru,
{\em Phys.\ Rev.}\ {\bf D58} (1998) 125006.
%
\bibitem{Anton98} F.~Antonuccio, O.~Lunin, H.-C.~Pauli,
S.~Pinsky, S.~Tsujimaru,
{\em Phys.\ Rev.}\ {\bf D58} (1998) 105024.
%
\bibitem{this work}
J.R. Hiller, O. Lunin, S. Pinsky, U. Trittmann,
{\em Phys.Lett.}~{\bf B482} (2000) 409--416.

%
\bibitem{Lanczos} C. Lanczos,
    {\em J. Res.\ Nat.\ Bur.\ Stand.}\ {\bf 45}, 255 (1950);
    J. Cullum and R.A. Willoughby,
    {\em Lanczos Algorithms for Large Symmetric Eigenvalue Computations},
    Vol.~I and II, (Birkhauser, Boston, 1985).
%
\bibitem{Haydock} R. Haydock, in {\em Solid State Physics}, Vol.~35,
 H. Ehrenreich, F. Seitz, and D. Turnbull (eds.),
(Academic, New York, 1980), p.~283.
%
\bibitem{osp93} H. Osborn and A.C. Petkou,
dimensions.
{\em Ann. Phys.}\ {\bf 231} (1994) 311.
%
\bibitem{Kutasov93} D. Kutasov,
{\em Phys.\ Rev.}\ {\bf D48} (1993) 4980.
%
\bibitem{bdk93} G. Bhanot, K. Demeterfi, and I.R. Klebanov,
{\em Phys.\ Rev.}\ {\bf D48} (1993) 4980.
%
\bibitem{hhlp99} P. Haney, J.R. Hiller, O. Lunin, S. Pinsky, and U.
Trittmann,
{\em Phys. Rev.}\ {\bf D62} (2000) 075002,  hep-th/9911243.
%
\bibitem{hpt01} J.R. Hiller, S. Pinsky, and U. Trittmann,
in preparation.
%
\end{chapthebibliography}
\end{document}